\documentclass[aps,prd,10pt,twocolumn,floatfix,altaffilletter,superscriptaddress,preprintnumbers,tightenlines,showpacs,showkeys,notitlepage,nofootinbib]{revtex4-2}

\usepackage{Preamble} 
\graphicspath{{images/}}
\usepackage{subfiles} 
\usepackage{placeins, changepage, comment}

\makeatletter
\renewcommand\onecolumngrid{%
  \do@columngrid{one}{\@ne}%
  \def\set@footnotewidth{\onecolumngrid}%
  \def\footnoterule{\kern-6pt\hrule width 1.5in\kern6pt}%
}
\makeatother

\definecolor{Red}{rgb}{1.,0.,0.}

\usepackage{fix-cm}

\begin{document}

\preprint{  MIT-CTP/6071}

\title{Higgs Couplings at a Future Wakefield Collider}

\author{Katherine Fraser}
\email{kfraser@berkeley.edu}
\affiliation{Leinweber Institute for Theoretical Physics, Department of Physics, University of California, \mbox{Berkeley}, CA 94720, USA}
\affiliation{Theoretical Physics Group, Lawrence Berkeley National Laboratory, Berkeley, CA 94720, USA}

\author{Simon Knapen}
\email{smknapen@lbl.gov}
\affiliation{Leinweber Institute for Theoretical Physics, Department of Physics, University of California, \mbox{Berkeley}, CA 94720, USA}
\affiliation{Theoretical Physics Group, Lawrence Berkeley National Laboratory, Berkeley, CA 94720, USA}

\author{Kevin Langhoff}
\email{langhoff@mit.edu}
\affiliation{Center for Theoretical Physics -- a Leinweber Institute, Massachusetts Institute of Technology,\\
77 Massachusetts Avenue, Cambridge, MA 02139, USA}

\author{Robert Szafron}
\email{rszafron@bnl.gov}
\affiliation{Department of Physics, Brookhaven National Laboratory, Upton, New York, 11973, U.S.A.}

\begin{abstract} 
We explore the potential of multiple possible future 10 TeV wakefield colliders to measure electroweak couplings of the Higgs boson. 
We find that the beam-beam interactions are not an impediment to high precision measurements of the Higgs couplings, provided that the luminosity spectra can be measured or calculated to high accuracy.
In addition to $e^+ e^-$ colliders, we also assess the effectiveness of alternatives such as $e^- e^-$ colliders or $\gamma \gamma$ colliders, which by-pass the positron acceleration challenge for wakefield colliders.
We find that a 10 $\text{ab}^{-1}$ dataset at a $\gamma\gamma$ collider yields qualitatively similar sensitivity to 10 $\text{ab}^{-1}$ at a muon collider and 1 $\text{ab}^{-1}$ at an $e^+e^-$ wakefield collider. 

\end{abstract}

\maketitle


\section{Introduction}

Extending the energy reach of collider experiments is essential for addressing several outstanding questions in particle physics.
Motivated by this goal, several proposals target a 10 TeV parton center-of-momentum (pCM) collider~\cite{Adolphsen:2022ibf, P5:2023wyd}. One promising approach envisions using linear plasma wakefield accelerators \cite{Verra:2026rvs,Gessner:2025acq}. This technology exploits strong electric fields ($\sim 10$--$100\;\text{GV/m}$) in laser- or beam-driven plasma wakes to accelerate charged particles over short distances. A 10 TeV center-of-mass collider based on this technology may fit within a facility of the order of several kilometers in length, compared to hundreds of kilometers for RF-based designs~\cite{Gessner:2025acq}. 
If competitive luminosity can be reached, a wakefield collider would combine an energy-frontier program with precision measurements in a single compact facility.

In this study, we consider two of the key technological considerations that currently shape the R\&D path towards a high-energy wakefield collider (WFC): the difficulty of maintaining adequate beam quality for positron beams and the challenge of preserving flat-beam geometries in plasma acceleration stages. Positron acceleration remains challenging for plasma wakefield accelerators because the plasma response does not naturally provide the stable, uniform focusing fields needed for high-quality positron beams \cite{Cao:2023ksg}. Flat beams, meanwhile, suffer from resonant emittance growth in plasma wakefield accelerators~\cite{Diederichs:2024zir}.
The latter is important because of the very high charge densities involved, which result in strong electro-magnetic interactions between the colliding beams. 
These beam-beam interactions generate hard radiation and $e^+e^-$ pair production at the interaction point, severely smearing out the luminosity across the whole energy  \cite{Esberg:2014zia, RevModPhys.94.045001}.
This effect could be mitigated to a degree, if beams with a flat profile can be delivered to the interaction point.
Recent advances in simulation have produced detailed luminosity spectra for several 10~TeV WFC beam configurations~\cite{simulationpaper, Schroeder:2022xdu,acceleratorpaper, warpx,ariannatalk,jenstalk}. It is therefore both timely and important to quantify how the technological challenges of positron acceleration and preserving flat-beam geometry affect the physics reach of a WFC, including its Higgs physics program, as inputs to the future wakefield accelerator R\&D program. 
Concretely, the challenges associated with positron acceleration and flat-beam preservation can be by-passed entirely with alternative collider configurations, such as round beam profiles, $e^-e^-$ collisions, and/or $\gamma\gamma$ collisions. 
The latter can be realized by Compton back-scattering laser light off two incoming $e^-$ beams.
In this paper, we quantify the extent to which the lack of flat $e^+e^-$ beams would or would not degrade the precision with which one could constrain the Higgs potential.  

A potential 10~TeV wakefield collider should also be viewed in the broader context of other future collider designs, including $e^+e^-$ colliders such as the FCC and photon colliders such as the XCC. The Future Circular Collider (FCC) program offers a fundamentally different approach from a wakefield machine, with a staged program consisting of an $e^+e^-$ factory (FCC-ee) for precision Higgs and electroweak measurements, followed by an 84~TeV proton-proton collider (FCC-hh) for direct discovery reach~\cite{FCC:2025lpp}. 
This multi-decade, multi-machine program would deliver percent- and sub-percent-level Higgs coupling measurements \cite{Altmann:2025feg} together with substantial direct discovery reach. 
The required infrastructure, however, is on a different scale than a WFC---the FCC tunnel circumference is approximately $91\,\text{km}$, compared to the potentially kilometer-scale footprint of a wakefield machine~\cite{Gessner:2025acq}. 
On the other hand, the acceleration technologies required for the FCC-ee are already within reach, while a WFC would still require a lengthy and uncertain R\&D path before construction could be considered.
The FCC therefore provides a useful reference point for the Higgs precision and high-energy reach: for a more speculative machine to be worth contemplating, it should at a minimum have comparable sensitivity to the FCC program.  

A complementary comparison is provided by the XFEL Compton Collider (XCC), a recently proposed photon-collider concept that uses RF acceleration and converts X-ray free-electron laser pulses into high-energy photon beams colliding in the $\gamma\gamma$ mode~\cite{Barklow:2023ess}. Owing to the sharply peaked photon-energy spectrum achievable with XFEL technology, the XCC can perform Higgs measurements with sensitivities comparable to those of $e^+e^-$ Higgs factories operating at substantially higher center-of-mass energies, while requiring a smaller accelerator complex~\cite{Barklow:2023ess,Castelazo:2026iuu}. The concept also admits an upgrade to $\sqrt{s_{\gamma\gamma}}\simeq 380~\mathrm{GeV}$, where the $\gamma\gamma\to hh$ cross section reaches its maximum in the Standard Model.
The XCC would be a natural predecessor of a multi-TeV wakefield collider and therefore provides another useful comparison point: It is interesting to ask whether a high energy WFC could further improve upon the precision delivered by the XCC, especially when assessing the physics potential of positron-free $\gamma\gamma$ collisions.

This paper is organized as follows:
In  \cref{sec:setup} we define our theoretical framework and the colliders we will study; \cref{sec:Event Generation and Processing} describes our analysis and  event-generation framework. Sections~\ref{sec:Single_Higgs_Analysis} and~\ref{sec:Double_Higgs_Analysis} present the single- and double-Higgs analyses, from which we derive sensitivities to the anomalous Higgs couplings. Finally, in \cref{sec:Conclusion} we compare the performance of $e^+e^-$, $e^-e^-$, and $\gamma\gamma$ collider options, assess the impact of round- and flat-beam geometries, and place the resulting sensitivities in the context of other future collider proposals.
We defer further details regarding our numerical analysis and analytical calculations to appendices \ref{app:analysisdetails} and \ref{app:vbf_amplitudes} respectively.

\section{Setup and definitions\label{sec:setup}}

\subsection{Higgs Precision Physics}
\label{sec:higgsprecision}

A 10~TeV wakefield collider simultaneously advances the energy frontier for direct particle discovery and the precision frontier for Higgs physics. High-resolution measurements of Higgs interactions complement direct searches by probing energy scales well beyond the kinematic limit of the machine~\cite{Adolphsen:2022ibf, P5:2023wyd}. 
In particular, determining the Higgs self-coupling is essential for reconstructing the Higgs potential, testing the mechanism of electroweak symmetry breaking, and probing scenarios such as electroweak baryogenesis~\cite{Dawson:2013bba,Reichert:2017puo}. Deviations in the Higgs self-coupling can significantly exceed those in single-Higgs couplings, making Higgs-potential measurements a uniquely sensitive probe of heavy new physics even when other Higgs and electroweak observables are measured to be close to the Standard Model prediction~\cite{Durieux:2022hbu,Bosse:2026bdk}. If heavy new states evade direct detection, anomalous Higgs couplings and self-interactions provide indispensable indirect signatures of new physics~\cite{LHCHiggsCrossSectionWorkingGroup:2016ypw,Brivio:2017vri}. Conversely, if new particles are produced on-shell, precision Higgs measurements will be essential for determining their underlying nature and their role in electroweak symmetry breaking~\cite{Dawson:2013bba}.

Modifications in Higgs couplings to electroweak bosons are typically parameterized in the electroweak broken phase using strength modifiers
\begin{align}\label{eq:lagrangiankappa}
    \mathcal{L} \supset~ &\frac{h}{v}\left(2\kappa_W m_W^2 W^{+\mu}W^-_\mu + \kappa_Z m_Z^2 Z^{\mu}Z_\mu - \frac{\kappa_3}{2} m_h^2 h^2 \right) \notag \\
    + &\frac{h^2}{v^2}\left(\kappa_{W2} m_W^2 W^{+\mu}W^-_\mu + \frac{1}{2}\kappa_{Z2} m_Z^2 Z^{\mu}Z_\mu  \right) +\cdots
\end{align}
where $\kappa_i = 1$ in the Standard Model (SM) and $\cdots$ represent higher orders in $h$ that we do not consider here. In what follows, we will often work with $\Delta \kappa_i\equiv \kappa_i -1$, which describe the deviations from the SM couplings.
This formalism has the advantage of being rather general, though it comes with a few important theoretical subtleties, as explained below. It is widely used for LHC measurements and projections for future colliders~\cite{LHCHiggsCrossSectionWorkingGroup:2013rie,LHCHiggsCrossSectionWorkingGroup:2016ypw,Ghezzi:2015vva,Han:2020pif}; we therefore adopt it as well in order to facilitate self-consistent comparisons. 
Current LHC measurements constrain Higgs gauge couplings at the $\mathcal{O}(5\text{--}10\%)$ level, while the self-coupling remains only weakly bounded~\cite{CMS:2022dwd,ATLAS:2022vkf}. The HL-LHC is projected to reach $\Delta \kappa_3 / \kappa_3 \sim 30\%$~\cite{ATLAS:2025eii}.

The coupling parametrization in~\cref{eq:lagrangiankappa} is most naturally interpreted as a
restricted leading-order Higgs Effective Field Theory (HEFT) parametrization.
In HEFT the physical Higgs is treated as an electroweak singlet, and the
coefficients $\{\kappa_W,\kappa_Z,\kappa_{W2},\kappa_{Z2},\kappa_3\}$ may be
taken as independent low-energy parameters. This is more general than the
Standard Model Effective Theory (SMEFT) \cite{Falkowski:2019tft,Cohen:2020xca,Cohen:2021ucp}, where the Higgs is embedded in an $SU(2)_L$ doublet, and gauge invariance
correlates the single- and double-Higgs couplings to electroweak gauge bosons.
In the restricted SM-like rescaling limit of dimension-six SMEFT, this UV structure implies additional relations between
$\kappa_{V2}$ and $\kappa_V$, for $V=W,Z$, up to consistently retained higher-order corrections \cite{Buchmuller:1985jz,Brivio:2017vri}. Under this assumption, deviations from the SM encoded in~\cref{eq:lagrangiankappa} can be related to a subset of SMEFT operators. Using the notation of \cite{Brivio:2020onw}, the operators
\begingroup
\small
\begin{equation}
\begin{gathered}
\frac{C_{H\Box}}{\Lambda^2}(H^\dag H) \Box (H^\dag H), \; \frac{C_{HD}}{\Lambda^2}(D^\mu H^\dag H)(H^\dag D_\mu H), \\[1ex]
\frac{C_{H}^{S}}{\Lambda^2}(H^\dag H)^3,
\end{gathered}
\end{equation}
\endgroup
can be formally related to those in \cref{eq:lagrangiankappa} through the matching relations \cite{Domenech:2025gmn} 
\begin{equation}
\begin{split}
    \Delta \kappa_{W} &= \Delta \kappa_{Z} = \frac{v^2}{\Lambda^2}\Big(C_{H\Box} - \frac{1}{4} C_{HD}\Big)\\
    \Delta \kappa_{W2} &= \Delta \kappa_{Z2} = \frac{4v^2}{\Lambda^2}\Big(C_{H\Box} - \frac{1}{4} C_{HD}\Big)
    \\
    \Delta \kappa_3 &= \frac{v^2}{\Lambda^2}\Big(- \frac{2 v^2}{m_H^2} C^S_H + 3\big(C_{H\Box} - \frac{1}{4} C_{HD}\big)\Big).
\end{split}
\end{equation}
Generic SMEFT deformations also generate derivative, tensor, and four-fermion contact
interactions that are not included in~\cref{eq:lagrangiankappa}. HEFT is the appropriate
low-energy description for broad classes of Higgs sectors whose geometry cannot
be described by a convergent expansion around a linear doublet, including
composite-Higgs-like scenarios~\cite{Giudice:2007fh,Alonso:2016oah,Cohen:2020xca,Falkowski:2019tft}. 

The $\kappa$-framework should therefore be viewed here as an anomalous-coupling
benchmark rather than as a complete, systematically improvable EFT description
of all high-energy observables. This distinction is especially important for
di-Higgs production through vector boson fusion~(VBF). When
\begin{equation}
\delta \equiv \kappa_{V2}-\kappa_V^2 \neq 0 ,
\end{equation}
the SM cancellation in the amplitude for the scattering of longitudinally polarized vector bosons ($V_LV_L\to hh$) is spoiled, and the amplitude grows
parametrically with the center-of-mass energy as $\delta\times m_{hh}^2/v^2$, as shown explicitly in Appendix~\ref{app:vbf_amplitudes}. 
This implies perturbative unitarity violation at a coupling-dependent scale,
schematically
\begin{equation}
  \Lambda_U \sim \frac{\sqrt{8\pi}\,v}{\sqrt{|\delta|}} ,
\end{equation}
up to channel-dependent $\mathcal{O}(1)$ factors~\cite{Cohen:2021ucp}. Thus, the
sensitivity of $|\delta|\sim 0.01$ projected below corresponds to
$\Lambda_U=O(10~{\rm TeV})$, comparable to the largest hard scales probed by a
10 TeV collider. The unrestricted fit should therefore be interpreted as a
benchmark sensitivity to anomalous longitudinal electroweak dynamics. A more
conservative EFT interpretation would impose a dynamical validity cut on the
hard scale, for example, $m_{hh}<\Lambda_U$.

\begin{figure*}[t]
    \centering
    \includegraphics[width=1\linewidth]{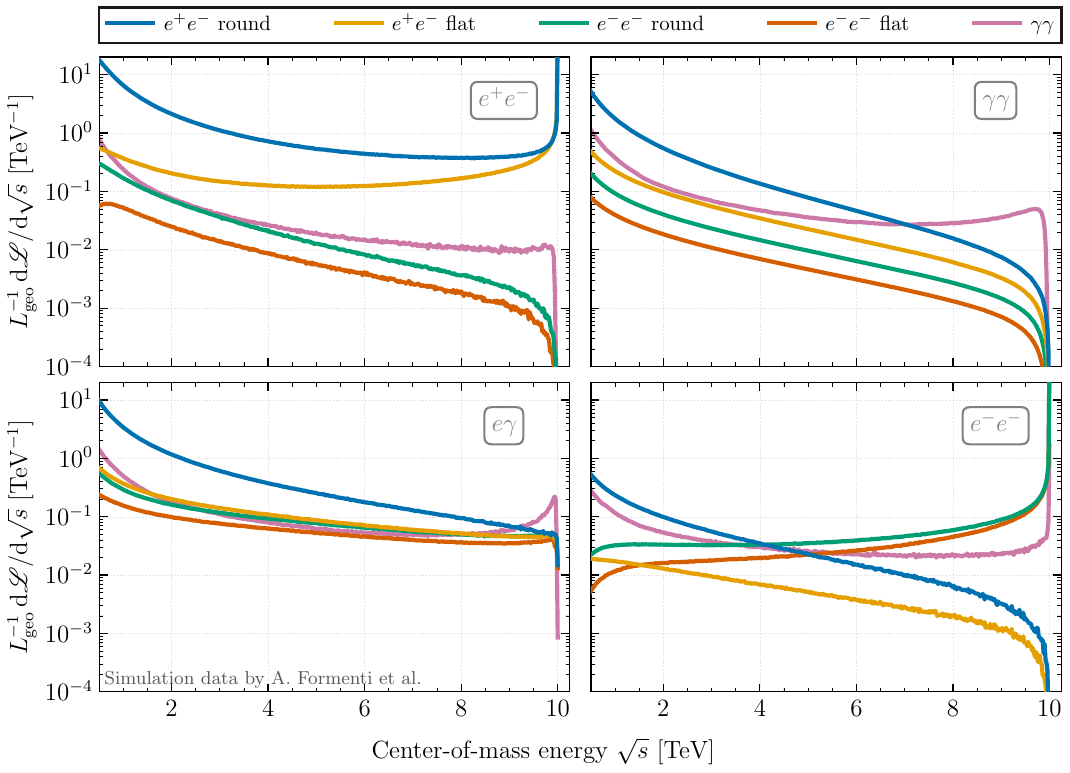}
    \caption{Luminosity spectra $\mathrm{d}\mathscr{L}/\mathrm{d}\sqrt{s}$ for 
    five WFC configurations for $e^+e^-$, $\gamma\gamma$, $e\gamma$, and $e^- e^-$ initial state collisions. Results are reproduced from \cite{simulationpaper,ariannatalk}, with beam parameters defined in \cite{Schroeder:2022xdu,acceleratorpaper,jenstalk}.  
    \label{fig:Lumi_Spectra}}
    
\end{figure*}

\subsection{Beam Dynamics and Collider Configurations}
\label{sec:colliders}
To arrive at competitive luminosities, wakefield colliders will need to deliver ultra-relativistic $\sim$nm size beams with charge as high as $\sim 10^9$ \cite{Schroeder:2022xdu,Gessner:2025acq,Verra:2026rvs}.
The resulting extreme charge densities 
lead to large beam-beam interactions, such as beamstrahlung and $e^+e^-$ pair production.
These effects spread the effective collision energy over a broad spectrum below the nominal $\sqrt{s}$ that is delivered at the interaction point (see \cref{fig:Lumi_Spectra} \cite{simulationpaper,ariannatalk}). 
Moreover, their importance depends strongly on the beam geometry: flat beams have asymmetric profiles in the transverse direction and generate less beamstrahlung than round beams with symmetric profiles  \cite{Schroeder_2022}.
A priori, beamstrahlung is a disadvantage for precision measurements, as it prevents one from precisely knowing the center-of-mass energy on a collision-by-collision basis.
Proposed linear RF accelerators such as the LCF and CLIC employ flat-beam designs for this reason.
Additionally, beamstrahlung reduces the luminosity available at the highest energies where VBF cross sections for Higgs production are largest. 
In wakefield accelerators, however, flat beams suffer from additional emittance growth \cite{Diederichs:2024zir}, and further R\&D advancements and/or an expanded beam delivery system would be needed to deliver such beams to the interaction area.
It is therefore important to quantify to what extent giving up on flat beams would affect the collider's physics potential. 
For all colliders we will assume that the systematic uncertainties on the luminosity spectra can be suppressed below the statistical uncertainty on the observable in question.

While beamstrahlung complicates the event-by-event determination of $\sqrt{s}$, the absence of a perfectly monochromatic center-of-mass energy is not intrinsically incompatible with precision measurements.
In particular, searches and precision measurements at the LHC routinely deal with a broad spectrum of partonic center-of-mass energies by carefully modeling the partonic luminosity spectrum and utilizing boost-invariant variables. 
Recent work has demonstrated that this also holds at high energy lepton colliders and even yields some advantages over mono-energetic collisions.
Concretely, for narrow resonances, the broad energy scan provided by beamstrahlung can improve sensitivity by orders of magnitude compared to monochromatic colliders~\cite{Cipressi:2026vector}. 
Electroweak pair production is improved in a similar manner~\cite{Chigusa:2025ewk}. 
Furthermore, VBF Higgs production, whose cross sections grow logarithmically with $s$, is expected to receive up to an order-of-magnitude enhancement in an $e^+e^-$ WFC from the increased total luminosity associated with beamstrahlung effects~\cite{Cipressi:2026vector}.

Another important consideration is that accelerating positrons is significantly more challenging than accelerating electrons with current wakefield accelerator technology \cite{PhysRevE.64.045501,Hogan:2003bs,Cao:2023ksg}. 
As the technology stands today, it is plausible that either positron beams cannot be realized in a future WFC or that the total geometric luminosity for collisions involving them will be significantly less than the luminosity achievable with electron beams. 
This strongly motivates the consideration of alternative beam configurations, such as collisions of only electron beams ($e^-e^-$), or collisions of photon beams ($\gamma\gamma$) generated via Compton backscattering a laser off incoming electron beams \cite{Asner:2001vh,Castelazo:2026iuu, Qureshi:2026ylt}.
For Higgs precision measurements, one may be tempted to assume  that any 10 TeV lepton collider can effectively be regarded as \emph{an electroweak boson collider}, thus rendering the electric charge of the initial state less relevant.
We test this hypothesis quantitatively and find the answer to be nuanced and dependent on the luminosity assumptions that are made for each collider (see \cref{sec:Conclusion}).

\subsection{Simulation Details}

We perform a comparison of Higgs coupling sensitivities for several 10 TeV WFC configurations. We consider $e^+e^-$ and $e^-e^-$ beam configurations with both round and flat beam geometries, as well as a $\gamma\gamma$ collider obtained by Compton back-scattering a laser off the $e^-$ beams of the $e^-e^-$ WFC with round beam geometry. The accelerator parameters are taken from \cite{Schroeder:2022xdu} and from a detailed optimization study by physicists in the LBNL BELLA group, to be published in a forthcoming paper \cite{acceleratorpaper}.
The luminosity spectra were simulated with \texttt{WarpX}~\cite{warpx} for $e^+e^-$ and $e^-e^-$ configurations, and \texttt{CAIN}~\cite{Chen:1994jt} for the $\gamma\gamma$ configuration by physicists in the LBNL Advanced Modeling Program \cite{simulationpaper,ariannatalk}. In \cref{fig:Lumi_Spectra} we reproduce their results for the initial states most relevant to our study.

\begin{figure*}[ht]
    \centering
    \includegraphics[width=1\linewidth]{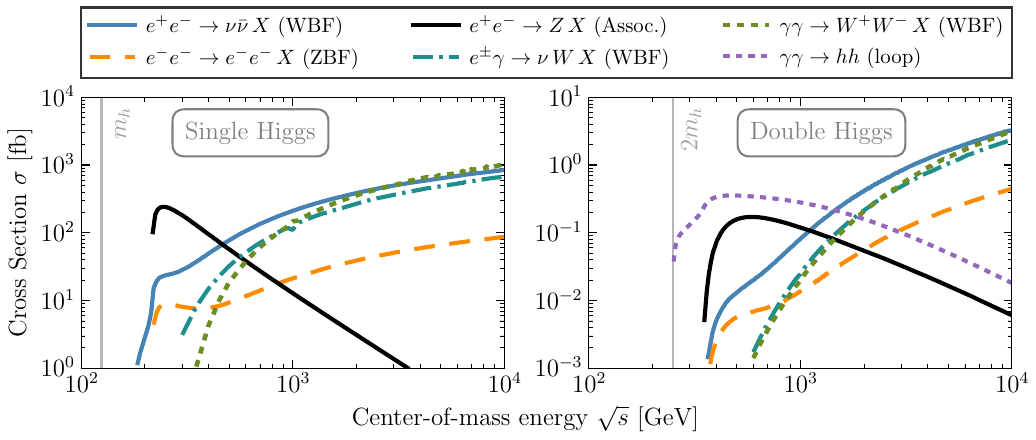}
    \caption{Cross sections vs center-of-mass energy $\sqrts$ for single Higgs (left) and double-Higgs (right) production processes, assuming the Standard Model values for the Higgs couplings. Line styles denote initial states: solid ($e^+e^-$), dashed ($e^-e^-$), dash-dotted ($e^\pm\gamma$), and dotted ($\gamma\gamma$). In the legend ``X'' stands for $h$ or $hh$, depending on the channel. }
    \label{fig:Cross_Sections}
\end{figure*}

All configurations assume $N_e = 1.2\times 10^9$ $e^\pm$/bunch and a longitudinal bunch size of $\sigma_z = \SI{8.5}{\micro\meter}$.  The flat and round beam configurations respectively assume transverse bunch sizes of $(\sigma_x,\sigma_y) = (6,\,0.4)\,\SI{}{\nano\meter}$ and $(\sigma_x,\sigma_y) = (1.55,\,1.55)\,\SI{}{\nano\meter}$. 
Throughout, we will assume a total geometric luminosity of $10\text{ ab}^{-1}$.\footnote{The geometric luminosity is the luminosity without accounting for beam-beam interactions and thus a measure of the luminosity delivered by the accelerator to the interaction region. It is defined by
$   \mathcal{L}_{geo} \equiv N_e^2 \nu_\text{rep} T/4\pi \sigma_x \sigma_y$
where $N_e$ is the number of electrons (or positrons) per bunch, $\nu_\text{rep}$ is the repetition rate, $\sigma_{x,y}$ are the transverse bunch sizes, and $T$ is the total run time of the collider.
The luminosity available for particle collisions differs from the geometric luminosity and is given by the curves in \cref{fig:Lumi_Spectra}.
} 
With a 10 kHz repetition rate, this would correspond to a run time of $\sim 6.5$ years.

To quantify how the beam-beam interactions modify the sensitivity, we will also analyze a hypothetical, mono-energetic $\ell^+\ell^-$ collider at 10 TeV with a luminosity of $10\,\text{ab}^{-1}$. 
We label this configuration as ``MuC'' in our results below, as it is also a good proxy for a 10 TeV muon collider with a luminosity of $10\,\text{ab}^{-1}$. Our results for this machine reproduce the muon collider sensitivity study by Han et.al.~\cite{Han:2020pif}, and therefore also serve as a cross check of our analysis pipeline.

\section{Event Generation and Processing}
\label{sec:Event Generation and Processing}

In this section, we discuss Higgs boson production via the dominant production channel, vector boson fusion. In a WFC, we can take advantage of all $e^\pm$ and $\gamma$ particles in the collisions to produce Higgs bosons, including $e^\pm \gamma$, $e^-e^-$, and $\gamma\gamma$ collisions. The production processes we consider are listed in \cref{tab:Production Modes}. 

\begin{table}[t] 
\begin{ruledtabular}
\begin{tabular}{lcl}
Process  & Main Process \\ \hline \hline
$e^+e^- \to \nu_e\bar{\nu}_e \, X$  & W-boson fusion (WBF) \\
$e^+e^- \to e^+e^- \, X$  & Z-boson fusion (ZBF)\\
$e^+e^- \to Z \, X$  & Associated Production\\
\hline
$e^-e^- \to e^-e^- \, X$    & Z-boson fusion (ZBF)      \\
\hline
$e^\pm\gamma \to \nu_e \, W^\pm X$ & $\gamma e$-induced WBF ($\gamma e$-WBF) \\
\hline
$\gamma \gamma \to W^+\, W^- X$  & $\gamma \gamma$-induced WBF ($\gamma\gamma$-WBF)  \\
$\gamma \gamma \to X$  &   photon fusion
\end{tabular}
\end{ruledtabular}
\caption{Signal processes where $X = h$
or $X = hh$.}
\label{tab:Production Modes}
\end{table}

We consider searches for Higgs bosons decaying into $b\bar{b}$ pairs only; subdominant decay channels, such as $h\to WW^\ast$, are not expected to qualitatively enhance the sensitivity to the triple Higgs coupling at a high energy lepton collider \cite{Roloff:2019crr}. The dominant backgrounds, therefore, come from continuum $b\bar b$ production and events where one or more $Z$ bosons are misidentified as Higgs bosons.
All signal and background events are generated with \texttt{MadGraph5\_aMC@NLO}~\cite{Alwall:2014hca}. 
The inclusive cross sections as a function of the center-of-mass energy of the colliding particles are shown in \cref{fig:Cross_Sections}, assuming the Standard Model ($\Delta \kappa_i =0$). 
We see that Z boson fusion (ZBF) is suppressed compared to W boson fusion (WBF), because the $Z$-lepton couplings in the Standard Model happen to be substantially smaller than the $W$-lepton couplings.
All electroweak bosons are decayed to quarks and leptons. We subsequently reweight all events to account for the branching ratios, luminosity spectra of the various colliders, acceptance cuts, and the dependence of the cross sections on the Lagrangian parameters in \eqref{eq:lagrangiankappa}. This procedure is described in detail in \cref{app:reweight}.

We do not perform a parton shower plus detector simulation but add a gaussian smearing to the truth-level b-quark momenta to model the jet energy resolution: 
for the jet energy resolution, we assume $\delta E/E$ of 10\%, along with a negligible uncertainty in the angular components of the jet momenta. 
On the $Z$-pole, this translates into an uncertainty on the invariant mass of the $b$-quarks ($m_{bb}$) of $\delta m_{bb}/m_{bb}\approx 7.5\%$.
This resolution determines to what extent $Z$ bosons can be separated from Higgs bosons and thus informs the background rates. 
For reference, ATLAS has achieved a $\sim 10\%$ resolution on $m_{bb}$ at the Higgs pole in run 2 data \cite{ATLAS:2017cen}. 
Moreover, CLIC is projecting a resolution of $\sim 7\%$ at the $Z$-pole for its 3 TeV run \cite{CLICdp:2018vnx}, while accounting for the expected beam-induced $\gamma\gamma\to$ hadrons background. 
This background is expected to be larger at a 10 TeV WFC, especially with round beams. Folding the $\gamma\gamma\to$ hadrons rate \cite{Barklow:1443518} against the luminosity spectra in the upper right panel of \cref{fig:Lumi_Spectra}, we find a collision environment with an $\mathcal{O}(1-10)$ times more soft tracks than in CLIC, depending on the collider considered. 
In sum, we effectively assume that future analysis and detector advances will offset these higher beam-induced backgrounds and will deliver a resolution on $m_{bb}$ somewhat better than that at the LHC and comparable to the projected resolution at CLIC. Ongoing work by members of the LBNL ATLAS group is expected to validate this assumption \cite{angiratalk}.

To suppress the various backgrounds, which contain $Z\to b\bar b$ decays, we will require the invariant mass of the two b-jets to satisfy $|m_{bb}-m_h|<15$ GeV. Assuming a gaussian energy resolution, the probability of a $Z\to b\bar b$ decay to pass this cut is then given by
\begin{align}
\frac{1}{2}\left[1 -\text{Erf}\left(\frac{110\,\text{GeV}-m_Z}{\sqrt{2}m_Z \times \delta m_{bb}/m_{bb}}\right)\right]\approx 0.003 \label{eq:Zfakerate}
\end{align}
with $\text{Erf}$ the error function and $\delta m_{bb}/m_{bb} =7.5\%$, as explained above. The expression in \eqref{eq:Zfakerate} merely serves as a rule-of-thumb; in our simulations, we perform the smearing directly on the $b$-quark momenta and subsequently impose all analysis cuts.
We have verified that increasing the mass resolution to $\delta m_{bb}/m_{bb}\approx 10\%$ does not change the results, as the backgrounds from $Z\to b\bar b$ remain negligible compared to the signal.

To maximize the signal efficiency, we only require one $b$-tag per $h$ candidate and assume an aggressive $b$-tagger with 90\% efficiency. 
For such a benchmark point, ATLAS has achieved roughly a 50\% mistag rate for $c$-jets and a 10\% mistag rate for light flavors \cite{ATLAS:2022qxm}. 
A similar performance is currently expected for CLIC \cite{CLICdp:2018vnx}, though it is plausible that this can be further improved  with recent machine learning methods.  If we adopt a $50\%$ charm mistag rate, this implies that about half of $Z\to c\bar c$ events would be misidentified as $Z\to b\bar b$. We account for this by rescaling the backgrounds with this correction factor. The contribution from mistagged light flavor jets is negligible.

We have assumed that the angular acceptance of the detector is $10^\circ < \theta < 170^\circ$, which corresponds to $|\eta|<2.44$ in units of pseudorapidity. This is roughly comparable to the acceptance of the phase I trackers of ATLAS and CMS. On the one hand, one may expect that the radiation environment ought to be less challenging than at a hadron collider, and it may therefore be feasible to extend this coverage. 
On the other hand, it is not yet known what constraints the beam delivery system would impose on the detector geometry. This is especially true for the $\gamma\gamma$ option, which would need laser systems in the forward region to be integrated into the detector design.

\section{Single Higgs Analysis}
\label{sec:Single_Higgs_Analysis}

\begin{table}[t]
    \begin{ruledtabular}
    \begin{tabular}{lc}
    Cut & Value \\
    \hline
    $b$-jet pseudorapidity    & $|\eta| < 2.44$ \\
    $b$-jet transverse momentum         & $\pT > 30\GeV$ \\
    Jet separation              & $\DeltaR_{bb} > 0.4$ \\
    Higgs mass window           & $|m_{bb} - 125\GeV| < 15\GeV$ \\
    $b\bar b$ transverse momentum                    & $30\,\text{GeV}<p_{T,bb} <500\,\text{GeV}$ \\
    \end{tabular}
    \end{ruledtabular}
    \caption{Event selection cuts for the single Higgs analysis. }   
    \label{tab:cutssingleH}
\end{table}

For the single Higgs analysis, we put minimal $p_T$, $\Delta R$, $m_{bb}$, and pseudorapidity cuts on the b-jets, as well as a cut on the vector sum of the transverse $b$-momenta, $p_{T,bb}\equiv|\vec p_{T,1}+\vec p_{T,2}|$ with $\vec p_{T,i}$ the transverse components of $b$-momentum vectors (see \cref{tab:cutssingleH}). The signal has high efficiency for these cuts, and the $p_{T,bb}$ in particular eliminates the otherwise large continuum backgrounds from the $e^+e^- \to b\bar b$ and $\gamma\gamma \to b\bar b$ processes. After these cuts, the dominant continuum backgrounds are from $e^+e^- \to b\bar b\gamma$ and $\gamma\gamma \to b\bar b\gamma$. All signal and background yields, before and after cuts, are shown in \cref{tab:yields_single_higgs}. 
A priori, even the mono-energetic (``MuC'') collider receives contributions from $\gamma\gamma$ and $\ell\gamma$ arising from the electroweak structure of the incoming leptons~\cite{Dawson:1984gx,Han:2020uid,Garosi:2023bvq}. 
These contributions are, however, negligible compared to the VBF rate, and are therefore not included for simplicity. 

\begin{table*}[t!]
  \begin{ruledtabular}
    \begin{tabular}{l rrrrr r}
      Single Higgs  {\footnotesize($\times 10^6$)} & $e^+e^-$ round & $e^+e^-$ flat & $e^-e^-$ round & $e^-e^-$ flat & $\gamma\gamma$ & MuC$^a$ \\
      \hline
      \\[-6pt]
      $e^+e^- \to \nu\bar\nu h$
      & \textbf{32\,/\,13}
      & \textbf{9.5\,/\,3.8}
      & \textbf{0.37\,/\,0.14}
      & 0.13\,/\,0.052
      & \textbf{0.55\,/\,0.21}
      & \textbf{5.0\,/\,1.9}
      \\[2pt]
      $ee \to ee\, h$
      & 3.5\,/\,1.5
      & 0.98\,/\,0.41
      & \textbf{0.33\,/\,0.14}
      & \textbf{0.32\,/\,0.14}
      & 0.12\,/\,0.051
      & 0.51\,/\,0.21
      \\[2pt]
      $\gamma\gamma \to W^+W^- h$
      & \textbf{4.5\,/\,2.4}
      & 0.92\,/\,0.48
      & \textbf{0.39\,/\,0.20}
      & \textbf{0.18\,/\,0.094}
      & \textbf{1.8\,/\,0.85}
      & ---
      \\[2pt]
      $e\gamma \to W\nu h$
      & \textbf{8.8\,/\,4.1}
      & \textbf{2.2\,/\,1.0}
      & \textbf{1.0\,/\,0.47}
      & \textbf{0.69\,/\,0.32}
      & \textbf{1.1\,/\,0.46}
      & ---
      \\[2pt]
       $\gamma\gamma \to H$$^b$
      & 1.5\,/\,-
      & 0.069\,/\,-
      & 0.029\,/\,-
      & 0.010\,/\,-
      & 0.28\,/\,-
      & ---
      \\[2pt]
      $e^+e^- \to Zh$$^b$
      & 7.5 / -
      & 0.17 / -
      & $< 0.01$ / -
      & $< 0.01$ / -
      & 0.11 / -
      & ---
      \\[4pt]
      \hline
      \\[-6pt]
      Total Signal
      & 58\,/\,21
      & 14\,/\,5.7
      & 2.1\,/\,0.96
      & 1.3\,/\,0.61
      & 4.0\,/\,1.6
      & 5.5\,/\,2.1
      \\[4pt]
      \hline \hline
      Backgrounds {\footnotesize($\times 10^6$)}~~ & $e^+e^-$ round & $e^+e^-$ flat & $e^-e^-$ round & $e^-e^-$ flat & $\gamma\gamma$ & MuC$^a$ \\
      \hline
      \\[-6pt]
      $e^+e^-,\,\gamma\gamma \to b\bar b \gamma$
      & \textbf{61\,/\,2.4}
      & 0.95\,/\,0.038
      & 0.24\,/\,0.010
      & 0.026\,/\,0.0011
      & 1.9\,/\,0.073
      & ---
      \\[2pt]
      $e\gamma \to e\, b\bar b$
      & 480\,/\,0.069
      & 58\,/\,0.0038
      & 26\,/\,0.00086
      & 15\,/\,0.00033
      & 42\,/\,0.0042
      & ---
      \\[2pt]
      $e\gamma \to W\nu\, b\bar b$
      & 12\,/\,0.070
      & 3.8\,/\,0.0058
      & 1.8\,/\,0.0016
      & 1.3\,/\,0.00074
      & 2.2\,/\,0.0057
      & ---
      \\[2pt]
      $\gamma\gamma \to W^+W^-\, b\bar b$
      & 39\,/\,1.3
      & \textbf{5.9\,/\,0.13}
      & \textbf{2.5\,/\,0.054}
      & \textbf{1.1\,/\,0.022}
      & \textbf{14\,/\,0.18}
      & ---
      \\[2pt]
      $e^+e^- \to \nu\bar\nu Z$
      & 38\,/\,0.019
      & 11\,/\,0.0045
      & 0.43\,/\,0.00019
      & 0.15\,/\,0.000070
      & 0.67\,/\,0.00031
      & \textbf{5.7\,/\,0.0042}
      \\[2pt]
      $ee \to ee\, Z$
      & 59\,/\,0.033
      & 4.0\,/\,0.00085
      & 0.91\,/\,0.000065
      & 0.79\,/\,0.000030
      & 0.99\,/\,0.00024
      & 1.7\,/\,0.000017
      \\[2pt]
      $\gamma\gamma \to W^+W^- Z$
      & 9.5\,/\,0.0056
      & 1.8\,/\,0.0010
      & 0.74\,/\,0.00043
      & 0.34\,/\,0.00020
      & 3.3\,/\,0.0019
      & ---
      \\[2pt]
      $e\gamma \to W^-\nu\, Z$
      & 12\,/\,0.0069
      & 3.0\,/\,0.0017
      & 1.4\,/\,0.00074
      & 0.91\,/\,0.00051
      & 1.5\,/\,0.00082
      & ---
      \\[2pt]
      $e\gamma \to e\,Z$
      & 59\,/\,0.018
      & 2.8\,/\,0.00069
      & 1.3\,/\,0.00022
      & 0.47\,/\,0.000087
      & 4.3\,/\,0.00027
      & ---
      \\[4pt]
      \hline
      \\[-6pt]
      Total Background
      & 770\,/\,3.9
      & 91\,/\,0.19
      & 35\,/\,0.068
      & 20\,/\,0.025
      & 71\,/\,0.26
      & 7.4\,/\,0.0042
      \\[2pt]
      $S/B$ (uncut\,/\,cut)
      & 0.075\,/\,5.4
      & 0.15\,/\,30
      & 0.060\,/\,14
      & 0.065\,/\,24
      & 0.056\,/\,6.2
      & 0.74\,/\,500
    \end{tabular}
\end{ruledtabular}
\begin{flushleft}
    {\footnotesize
      $^a$MuC: for muon collider replace $e \leftrightarrow \mu$ in each row label. \\
      $^b$Before cuts only. After cuts rate is negligible compared to VBF and not included in the analysis.}
  \end{flushleft}
  \caption{Event yields assuming a geometric luminosity of $10~\iab$ for single Higgs production ($h \to b\bar{b}$) and its backgrounds (in millions of events). Branching ratios to $b$-quarks are included in event counts. Each cell shows Uncut\,/\,Cut, where ``Cut'' applies the acceptance and selection criteria of Table~\ref{tab:cutssingleH}. \textbf{Bold} entries contribute $\geq 10\%$ of the total cut yield at that collider. 
  }
  \label{tab:yields_single_higgs}
\end{table*}

In addition to VBF, the Higgs boson can be produced resonantly from beamstrahlung photons with $E_{\rm cm}\approx m_h$ through the loop-induced $H\gamma\gamma$ interaction.\footnote{We assume that all deviations of the $h\gamma\gamma$ interactions come from the $\kappa_{W,Z}$ dependence in the loop diagrams. This assumption needs to be generalized if there are new heavy charged particles with a coupling to $h$. However, one would generally expect to first discover such particles directly in the colliders we consider here \cite{Chigusa:2025ewk}.} In the narrow-width approximation, the number of events is
\begin{align}
    N(\gamma \gamma \to h) = \frac{2\pi^2 \Gamma(h\to \gamma \gamma)}{ m_h^2} \frac{d \mathscr{L}_{\gamma \gamma}}{d \sqrt{s}}\bigg|_{\sqrt{s} = m_h}\,,
\end{align}
with $d\mathscr{L}_{\gamma \gamma}/d \sqrt{s}\big|_{\sqrt{s} = m_h}$ the  luminosity spectrum summed over all photon polarizations.
This channel was shown to be promising at lower-energy photon-photon colliders~\cite{Berger:2025ijd,Castelazo:2026iuu,Qureshi:2026ylt}. At $10\TeV$, the resonant rate before cuts is comparable to VBF, but the two beamstrahlung photons are collinear and the Higgs is produced at rest, so $p_{T,bb}\simeq0$. The $p_{T,bb}>30\GeV$ requirement of \cref{tab:cutssingleH}, imposed to suppress the $b\bar b$ continuum, therefore removes the resonant signal as well, leaving a negligible contribution after selection.
Associated production ($e^+e^- \to Zh$) is also negligible.

We can then write the single Higgs production cross section as
\begin{equation}
  \label{eq:single_higgs_xsec}
  \sigma_{VBF}(\kW,\kZ) \;=\; \kW^2\,\sigma_{WBF}^{\rm SM} \;+\; \kZ^2\,\sigma_{ZBF}^{\rm SM}\,,
\end{equation}
where ``$\sigma_{WBF}^{\rm SM}$'' represents the sum of the $e^+ e^-\to \nu_e \bar{\nu}_e\,h$, $e^-\gamma\to\nu_e W^-\,h$, and $\gamma \gamma\to W^+ W^-\,h$ processes, and ``$\sigma_{ZBF}^{\rm SM}$'' represents $ee\to ee\,h$.
This provides a clean handle on $\kW$ and $\kZ$ if both classes of processes can be measured separately. 
The relative weight of the two channels and the degree to which they can be separated depend strongly on the beam configuration.

\begin{figure}[t]
    \centering
    \includegraphics[width=1\linewidth]{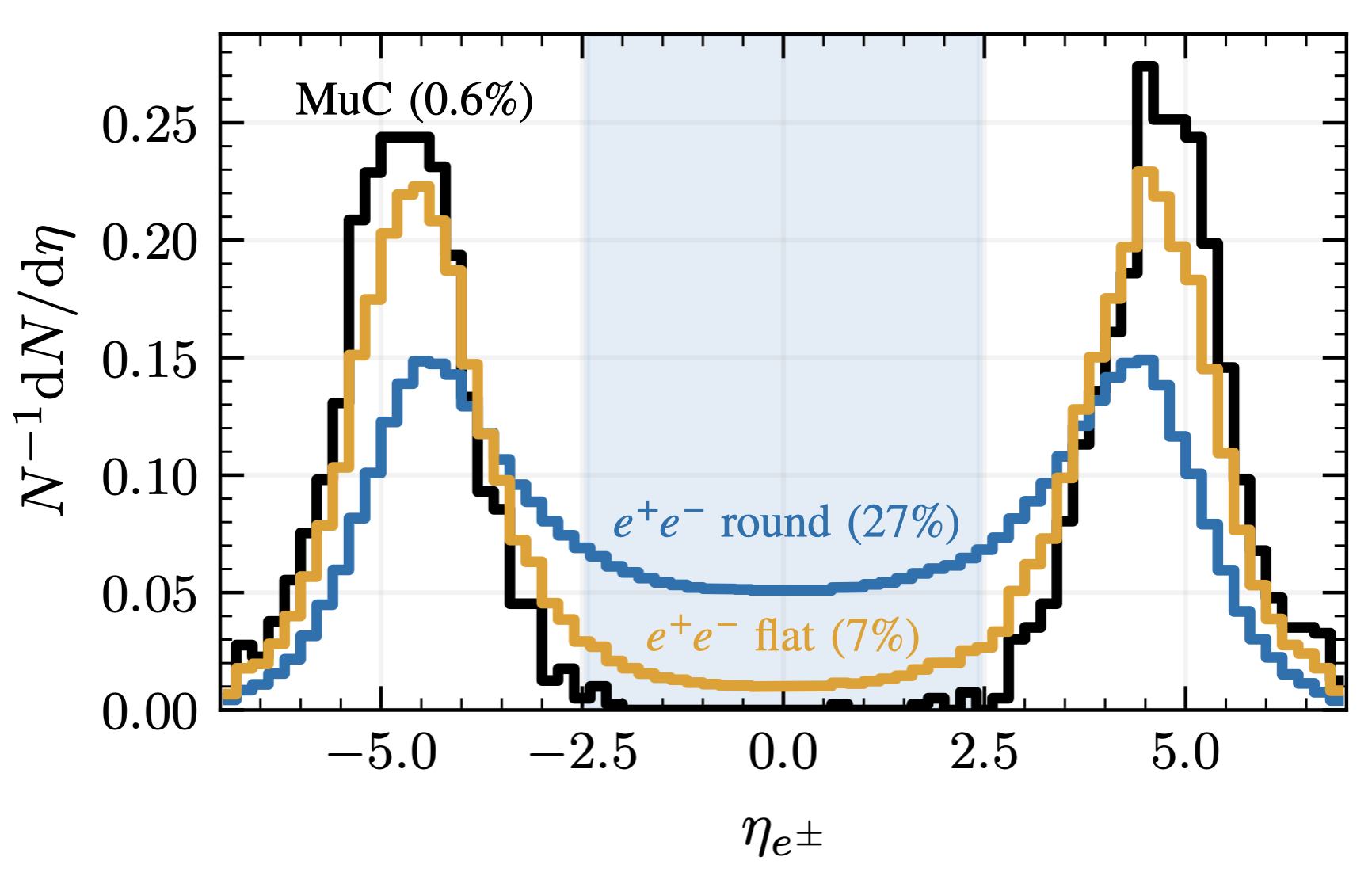}
    \caption{Normalized electron pseudorapidity distributions and fraction of events with at least one charged lepton in $|\eta|<2.44$ for ZBF ($\ell^+\ell^-\to \ell^+\ell^- H$) with $e^+ e^-$ round and flat beams and a 10\TeV\ \MuC. Low energy beamstrahlung-induced collisions yield more central electrons, improving tagging efficiency and ZBF distinguishability.}
    \label{fig:Rap_Eff}
\end{figure}

We can single out the ZBF fraction by tagging a final-state electron with $|\eta_e|<2.44$; the $e$-tag also receives a $W\to e\nu$ contribution from WBF, which the likelihood accounts for. Fig.~\ref{fig:Rap_Eff} shows the rapidity distribution of the electrons, as well as the efficiency for $|\eta_e|<2.44$.
We see that round beams yield a higher efficiency, as more collisions take place at lower energy, thus reducing the degree to which the final state electrons are pushed beyond the detector's angular acceptance. Qualitative improvements are possible if the angular coverage of the detector can be increased (see \cref{app:coverage}).

We split the selected $h\to b\bar b$ events into three mutually exclusive categories: an \emph{$e$-tag} with a final-state electron in $|\eta|<2.44$, a \emph{$j$-tag} with a hadronic $W\to jj$ in acceptance, and an \emph{untagged} remainder. Since a final-state $W$ can decay either to $jj$ or to $e\nu$, one event may present both a jet tag and an electron; events are assigned by the priority $j\text{-tag}>e\text{-tag}>\text{untagged}$, keeping the categories disjoint. The $j$-tag takes precedence because a hadronic $W\to jj$ cleanly identifies a WBF event; routing these to the $j$-tag reduces the WBF contamination of the $e$-tag and sharpens the separation of $\kW$ from $\kZ$. The expected number of events in category $a$ is
\begin{align}
  \label{eq:single_higgs_incl}
  n_{a}(\vec\kappa) = \frac{\mathrm{BR}_{b\bar b}(\vec\kappa)}{\mathrm{BR}_{b\bar b}^{\rm SM}}\left(\kW^2\,n_{a,\mathrm{WBF}}^{\rm SM} + \kZ^2\,n_{a,\mathrm{ZBF}}^{\rm SM}\right)+ n_{a,\text{bkg}}
\end{align}
with $n_{a,\mathrm{WBF}}^{\rm SM}$, $n_{a,\mathrm{ZBF}}^{\rm SM}$ and $n_{a,\text{bkg}}$ the SM event rates for the WBF processes, the ZBF process, and the backgrounds in \cref{tab:yields_single_higgs}. The prefactor
\begin{equation}
  \frac{\mathrm{BR}_{b\bar b}(\vec\kappa)}{\mathrm{BR}_{b\bar b}^{\rm SM}} = \left[1 + \mathrm{BR}_{WW}^{\rm SM}(\kW^2-1) + \mathrm{BR}_{ZZ}^{\rm SM}(\kZ^2-1)\right]^{-1},
\end{equation}
with $\mathrm{BR}_{WW}^{\rm SM}=0.215$ and $\mathrm{BR}_{ZZ}^{\rm SM}=0.026$, encodes the $\kappa$-dependence of the total Higgs width.

To extract constraints on the Higgs couplings, we employ a Poisson
log-likelihood of the form
\begin{equation}
\label{eq:logL}
-\Delta\ln\mathcal L(\vec\kappa)
=
\sum_a
\left[
n_a^{\rm pred}
-
n_a^{\rm obs}
+
n_a^{\rm obs}
\ln
\frac{n_a^{\rm obs}}{n_a^{\rm pred}}
\right].
\end{equation}
For the single-$h$ analysis, the index $a$ runs over the three tag categories (untagged, $e$-tag, $j$-tag), $n_a^{\rm obs}$ is the SM expectation, and the parameter vector is $\vec\kappa=(\kW,\kZ)$.
We do not include systematic uncertainties at this stage. 
For example, the luminosity spectra themselves are a possible source of such systematic uncertainties. It is unclear at this time whether they could be computed with the required precision. However, it is likely that they can be measured with high accuracy through high rate SM processes such as $e^+e^-\to e^+e^-$ and $e^\pm \gamma \to e^\pm \gamma$.
We assume that the uncertainties from such an extraction are small compared to the statistical uncertainties for the Higgs production rate itself.

\section{Double Higgs Analysis}
\label{sec:Double_Higgs_Analysis}

\begin{table}[t]
    \begin{ruledtabular}
    \begin{tabular}{lc}
    Cut & Value \\
    \hline
    $b$-jet pseudorapidity    & $|\eta| < 2.44$ \\
    $b$-jet transverse momentum         & $\pT > 30\GeV$ \\
    Jet separation              & $\DeltaR_{bb} > 0.4$ \\
    Higgs mass window           & $|m_{bb} - 125\GeV| < 15\GeV$ \\
    \end{tabular}
    \end{ruledtabular}
    \caption{Event selection cuts for the double-Higgs analysis. The $m_h$ window cut is applied to the b-jet pairing that minimizes $(m_{b_1b_2} - m_h)^2 +  (m_{b_3b_4} - m_h)^2$. 
   }   
    \label{tab:cutsdoublehiggs}
\end{table}

For double-Higgs production, the rate is provided by the various vector boson fusion channels as well as photon fusion (see \cref{fig:Cross_Sections}). 
The importance of each channel will depend strongly on the collider.
 The cross sections now depend on $\kthree$, $\kappa_{W}$, $\kappa_{Z}$, $\kappa_{W2}$ and $\kappa_{Z2}$. In what follows, we will set $\kappa_W=\kappa_Z=1$, effectively assuming that they have been measured at the SM value with high precision, either at a lower energy Higgs factory or through the procedure in \cref{sec:Single_Higgs_Analysis}. In addition, we assume no direct new physics contributions to $\gamma\gamma \to h h$, except for deviations in $\kappa_{W2}$ and $\kthree$.

\begin{table*}[t]
  \begin{ruledtabular}
    \begin{tabular}{l rrrrr r}
      Di-Higgs {\footnotesize($\times 10^3$)} & $e^+e^-$ round & $e^+e^-$ flat & $e^-e^-$ round & $e^-e^-$ flat & $\gamma\gamma$ & MuC$^a$ \\
      \hline
      \\[-6pt]
      $e^+e^- \to \nu\bar\nu hh$
      & \textbf{35\,/\,5.5}
      & \textbf{18\,/\,2.5}
      & 0.31\,/\,0.046
      & 0.13\,/\,0.019
      & 0.49\,/\,0.079
      & \textbf{11\,/\,1.5}
      \\[2pt]
      $ee \to ee\, hh$
      & 4.7\,/\,0.83
      & 2.4\,/\,0.36
      & \textbf{0.86\,/\,0.12}
      & \textbf{0.89\,/\,0.12}
      & 0.22\,/\,0.036
      & \textbf{1.5\,/\,0.21}
      \\[2pt]
      $\gamma\gamma \to hh$
      & \textbf{3.3\,/\,1.5}
      & 0.30\,/\,0.12
      & 0.13\,/\,0.051
      & 0.049\,/\,0.021
      & \textbf{0.71\,/\,0.15}
      & ---
      \\[2pt]
      $\gamma\gamma \to W^+W^- hh$
      & 2.4\,/\,0.63
      & 0.63\,/\,0.16
      & \textbf{0.27\,/\,0.068}
      & 0.13\,/\,0.032
      & \textbf{1.7\,/\,0.36}
      & ---
      \\[2pt]
      $e\gamma \to W\nu hh$
      & \textbf{6.6\,/\,1.4}
      & \textbf{2.5\,/\,0.48}
      & \textbf{1.2\,/\,0.22}
      & \textbf{0.83\,/\,0.16}
      & \textbf{1.3\,/\,0.22}
      & ---
      \\[2pt]
      $e^+e^- \to Zhh$$^b$
      & 5.3 / -
      & 0.31  / -
      & 0.043  / -
      & 0.027  / -
      & 0.099  / -
      & ---
      \\[4pt]
      \hline
      \\[-6pt]
      Total Signal
      & 57\,/\,9.8
      & 24\,/\,3.6
      & 2.8\,/\,0.51
      & 2.1\,/\,0.35
      & 4.5\,/\,0.85
      & 13\,/\,1.7
      \\[4pt]
      \hline \hline
      Background {\footnotesize($\times 10^3$)}~~ & $e^+e^-$ round & $e^+e^-$ flat & $e^-e^-$ round & $e^-e^-$ flat & $\gamma\gamma$ & MuC$^a$ \\
      \hline
      \\[-6pt]
      $e^+e^- \to \nu\bar\nu\, Zh$
      & 100\,/\,0.55
      & 49\,/\,0.19
      & 0.93\,/\,0.0055
      & 0.38\,/\,0.0021
      & 1.4\,/\,0.0085
      & 31\,/\,0.092
      \\[2pt]
      $ee \to ee\, Zh$
      & 3.1\,/\,0.021
      & 0.49\,/\,0.0011
      & 0.13\,/\,0.00016
      & 0.13\,/\,0.000084
      & 0.080\,/\,0.00030
      & 0.22\,/\,0.000046
      \\[2pt]
      $e^+e^- \to \nu\bar\nu\, ZZ$
      & 150\,/\,0.80
      & 73\,/\,0.31
      & 1.4\,/\,0.0078
      & 0.56\,/\,0.0031
      & 2.2\,/\,0.012
      & 46\,/\,0.17
      \\[2pt]
      $ee \to ee\, ZZ$
      & 5.0\,/\,0.024
      & 0.80\,/\,0.0023
      & 0.24\,/\,0.00066
      & 0.22\,/\,0.00059
      & 0.13\,/\,0.00044
      & 0.37\,/\,0.00057
      \\[4pt]
      \hline
      \\[-6pt]
      Total Background
      & 260\,/\,1.4
      & 120\,/\,0.51
      & 2.7\,/\,0.014
      & 1.3\,/\,0.0058
      & 3.8\,/\,0.021
      & 78\,/\,0.26
      \\[2pt]
      $S/B$ (uncut\,/\,cut)
      & 0.22\,/\,7.1
      & 0.19\,/\,7.1
      & 1.0\,/\,36
      & 1.6\,/\,60
      & 1.2\,/\,40
      & 0.17\,/\,6.4
    \end{tabular}
  \end{ruledtabular}
  \begin{flushleft} \footnotesize{$^a$MuC: replace $e \leftrightarrow \mu$ in each row label. \\ $^b$Higgsstrahlung; before cuts only. Negligible compared to VBF and not included in the analysis.}
  \end{flushleft}
   \caption{Event yields for a geometric luminosity of $10~\iab$ for di-Higgs production ($hh \to b\bar{b}b\bar{b}$) and its backgrounds (in thousands of events). Branching ratios to $b$-quarks are included in event counts. Each cell shows Uncut\,/\,Cut, where ``Cut'' applies the acceptance and selection criteria of Table~\ref{tab:cutsdoublehiggs}. \textbf{Bold} entries contribute $\geq 10\%$ of the total cut yield at that collider. The backgrounds contribute negligibly to the likelihood in all cases.}
  \label{tab:yields_double_higgs}
  \begin{flushleft}

  \end{flushleft}
\end{table*}

The di-Higgs analysis fits three independent parameters, $\vec\kappa=(\Delta\kappa_{W2},\Delta\kappa_{Z2},\Delta\kappa_3)$, holding $\kW=\kZ=1$ fixed (so $\mathrm{BR}(H\to b\bar b)$ stays at its SM value). In each $(\sqrt{s},m_{hh})$ bin the rate is a sum of WBF, ZBF, and $\gamma\gamma\to hh$ contributions; WBF and the $\gamma\gamma\to hh$ processes depend on $(\Delta\kappa_{W2},\Delta\kappa_3)$ and ZBF depends on $(\Delta\kappa_{Z2},\Delta\kappa_3)$, with no $\Delta\kappa_{W2}\Delta\kappa_{Z2}$ cross term. Each contribution is obtained by rescaling the Standard Model prediction with a quadratic ratio in the $(\Delta\kappa_{W2},\Delta\kappa_{Z2},\Delta\kappa_3)$, weighted by coefficients $r_i(\sqrt{s},m_{hh})$ that are determined with MadGraph (see appendices~\ref{app:reweight} and ~\ref{subsec:rcoeffs}). An analogous approach has been used previously in the context of future muon colliders \cite{Han:2020pif}.

While we simulate the full kinematics without relying on the effective vector approximation, it is nevertheless instructive to study the underlying $VV\to hh$ amplitudes, which provide useful insight into the qualitative features of VBF production. This is discussed in Appendix~\ref{app:vbf_amplitudes}.

The $\gamma\gamma \to hh$ process proceeds at one loop through top-quark and $W$-boson
triangle and box diagrams~\cite{Jikia:1992mt,Belusevic:2004pz,Bharucha:2020bhy}.
In particular,  the opposite-helicity amplitude
$\mathcal{M}_{+-}$ receives contributions only from box diagrams and is therefore
independent of $\kappa_3$, while all sensitivity to the Higgs self-coupling resides
in the same-helicity amplitude $\mathcal{M}_{++}$, which contains both triangle and box terms.
We evaluated the cross sections with a custom implementation of the one-loop
integrals using \textsc{LoopTools}~\cite{Hahn:1998yk}, validated against
\textsc{MadGraph5\_aMC@NLO}~\cite{Alwall:2014hca}; see Appendix~\ref{app:ggHH} for details. 
In the SM, the cross section peaks near $\sqrt{s}\approx 380$ GeV and drops for higher $\sqrt{s}$.

For the di-Higgs search, we apply the selection cuts listed in \cref{tab:cutsdoublehiggs}. To suppress the combinatorial background, we choose the $b$-jet pairing that minimizes 
\begin{equation}
    (m_{b_1b_2} - m_h)^2 +  (m_{b_3b_4} - m_h)^2,
\end{equation}
after smearing the $b$-momenta. 
The resulting signal and background rates are shown in \cref{tab:yields_double_higgs}, assuming a geometric luminosity of 10 $\text{ab}^{-1}$ for all colliders. After cuts, the dominant component of the $ZZ$ background is due to two pairings of $b$-quarks from unrelated $Z$-bosons, each accidentally reconstructing to an $m_{bb}$ that passes the mass window cut. 
It is possible to further suppress this background by implementing a veto on events for which any pair of $b$-jets has an invariant mass close to $m_Z$.   
It is however already irrelevant in the likelihood, and we therefore for simplicity did not impose this additional cut. 

For the double-Higgs analysis, we also use the Poisson log-likelihood~\eqref{eq:logL}, where the index $a$ now runs over analysis bins, i.e.~the three tag categories (untagged, $e$-tag, $j$-tag) times eight $m_{4b}$ bins, and $\Delta\vec\kappa=(\Delta\kappa_{W2},\Delta\kappa_{Z2},\Delta\kappa_3)$ with $\Delta\kW=\Delta\kZ=0$.
Contours are drawn at fixed $\Delta\chi^2$ from the SM point: $\Delta\chi^2 = 2.30$ for two-parameter regions and $1.00$ for one-parameter intervals (both 68\% C.L.).
In both cases the remaining parameters are profiled out, unless stated otherwise.
Analogous to the single-$h$ analysis, we split the events into three mutually exclusive categories: untagged, $e$-tagged, and $j$-tagged. The $j$-tag ($W\to jj$) is fed only by WBF, while the $e$-tag collects electrons from ZBF and from the leptonic $W\to e\nu$ decay of WBF; the differing tag compositions, with the per-bin $m_{4b}$ shapes, separate $\kappa_{W2}$ from $\kappa_{Z2}$. 

\section{Results and Conclusion}

\begin{figure*}[p]
    \centering
    \includegraphics[width=1\linewidth]{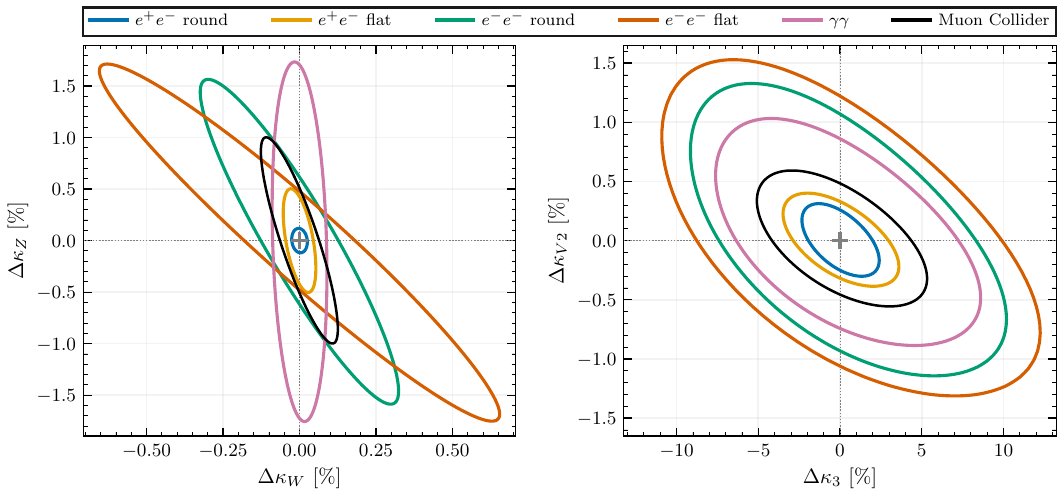}
    \caption{68\% C.L.\ contours at $10\,\mathrm{ab}^{-1}$ in the $(\Delta\kappa_W,\Delta\kappa_Z)$ plane (left) and the $(\Delta\kappa_3,\Delta\kappa_{V2})$ plane (right) for WFC configurations and a 10\TeV muon collider. In the right-hand panel we have assumed $\kappa_W=\kappa_Z=1$ and $\Delta\kappa_{V2} \equiv \Delta\kappa_{W2} = \Delta\kappa_{Z2}$.
    }
    \label{fig:Contours}
\end{figure*}

\begin{figure*}[p]
    \centering
    \includegraphics[width=1\linewidth]{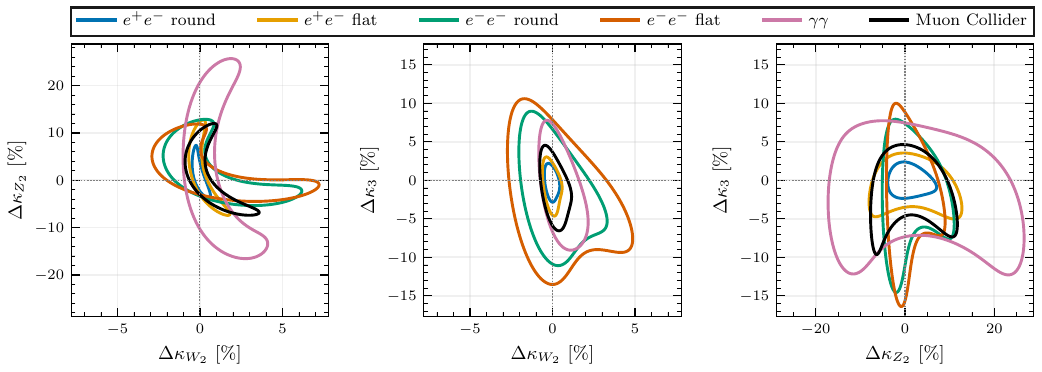}
    \caption{Projected 68\%~CL\ contours for di-Higgs sensitivity at $10\iab$ in 3D coupling space $(\Delta\kappa_{W2},\,\Delta\kappa_{Z2},\,\dkthree)$, with third coupling profiled while holding $\kW=\kZ=1$ fixed.} \label{fig:dh_corner_2x2}
\end{figure*}

\label{sec:Conclusion}

We now present the resulting Higgs coupling sensitivities and their dependence on the collider design. We begin by comparing the various wakefield collider configurations introduced in \cref{sec:colliders}, as shown in \cref{fig:Contours} and \cref{fig:dh_corner_2x2}. In \cref{fig:Contours} we fix $\Delta\kappa_{W2}=\Delta\kappa_{Z2}$, while in \cref{fig:dh_corner_2x2} we relax the $\Delta\kappa_{W2}=\Delta\kappa_{Z2}$ assumption and show the pairwise sensitivities in the full $(\Delta\kappa_{W2},\,\Delta\kappa_{Z2},\,\Delta\kappa_3)$ space, profiling the third coupling in each panel.

Starting with $\Delta \kappa_{W,Z}$, we see that $e^+e^-$ colliders achieve better precision on $\Delta\kappa_W$ because the WBF rate is strongly suppressed in the other colliders.
The ZBF rate is also somewhat larger, leading to a similar or somewhat better sensitivity on $\Delta\kappa_Z$ for the $e^+e^-$ flat beams, compared to the $e^-e^-$ and $\gamma\gamma$ colliders.
The $\Delta \kappa_Z$ precision achievable with the $e^+e^-$ round beams is still significantly better. This is in part because of the larger ZBF rate and in part due to the larger acceptance for the $e$-tag category (see \cref{fig:Rap_Eff}): Since more collisions occur at lower energies, the outgoing $e^{\pm}$ are less likely to be outside the detector's acceptance.
Both round and flat $e^+e^-$ colliders outperform the mono-energetic ``MuC'' benchmark with the same luminosity, indicating that beam-beam interactions in this context \emph{enhance} the sensitivity rather than hinder it.
This however assumes that any systematic uncertainties introduced by the beam-beam interactions will be controlled to a level better than the statistical uncertainty.
It seems plausible that this is possible, e.g.~by directly measuring the luminosity as a function of $\sqrt{s}$ through high rate QED processes such as $e^+e^-\to e^+e^-$ and $e^\pm\gamma\to e^\pm \gamma$. This has been studied in detail for the ILC and CLIC \cite{BozovicJelisavcic:2013lni,Poss:2013oea}, but to our knowledge, not yet for a high energy wakefield collider. This is therefore to be considered an assumption in our analysis, to be examined in future studies. 

Fixing $\kappa_{V2}=\kappa_{W2}=\kappa_{Z2}$, a fairly straightforward picture emerges in the right-hand panel of \cref{fig:Contours}, as the sensitivity is essentially set by the overall number of signal events in each collider (see \cref{tab:yields_double_higgs}). 
The $\Delta \kappa_{V2}$ coupling is most sensitive to the high $m_{hh}$ part of the signal spectrum, which explains the slight differences in shape between the ellipses for the different colliders. 
Figure~\ref{fig:dh_corner_2x2} shows a qualitatively similar picture in terms of the relative precision that can be achieved at each collider. The exception is $\Delta\kappa_{Z2}$ at the $\gamma\gamma$ collider, for which the expected precision is worse than that for the other colliders. As in the single-Higgs case, this behavior arises from the comparatively low ZBF rate at the $\gamma\gamma$ collider (see \cref{tab:yields_double_higgs}).

We note that the $e$-tag and $j$-tag categories are less relevant to the likelihood than for single Higgs production, and the sensitivity is primarily driven by the untagged category (see \cref{app:tag}). This is because $e$-tagging is efficient only at low $\sqrt{s}$, whereas the unitarity-violating effects of $\Delta\kappa_{Z2}\neq0$ grow with energy and dominate the high-$m_{4b}$ tail, where $e$-tagging is inefficient. The $\Delta\kappa_{Z2}$ sensitivity therefore mostly resides in the shape of the untagged high-mass spectrum and survives the removal of $e$-tagging, unlike in single-Higgs where the forward tag significantly aids in isolating the ZBF ($\Delta\kappa_Z$) contribution. 
The comparatively low sensitivity of the $e$ and $j$-tagged categories also creates degeneracies in the likelihood, especially between $\Delta\kappa_{W2}$ and $\Delta\kappa_{Z2}$: Effectively, the signal amplitude is a quadratic function of the $\Delta\kappa_{W2}$, $\Delta\kappa_{Z2}$, and $\Delta\kappa_{3}$, and a deviation in one parameter can be (partially) absorbed by a deviation in another. 
This degeneracy causes the extended contour shapes in e.g.~the $\Delta\kappa_{Z2}$ direction. The concave contours in \cref{fig:dh_corner_2x2} indicate that the non-linear terms in the likelihood have an $\mathcal{O}(1)$ effect.

\begin{figure*}[!t]
    \centering
    \includegraphics[width=\linewidth]{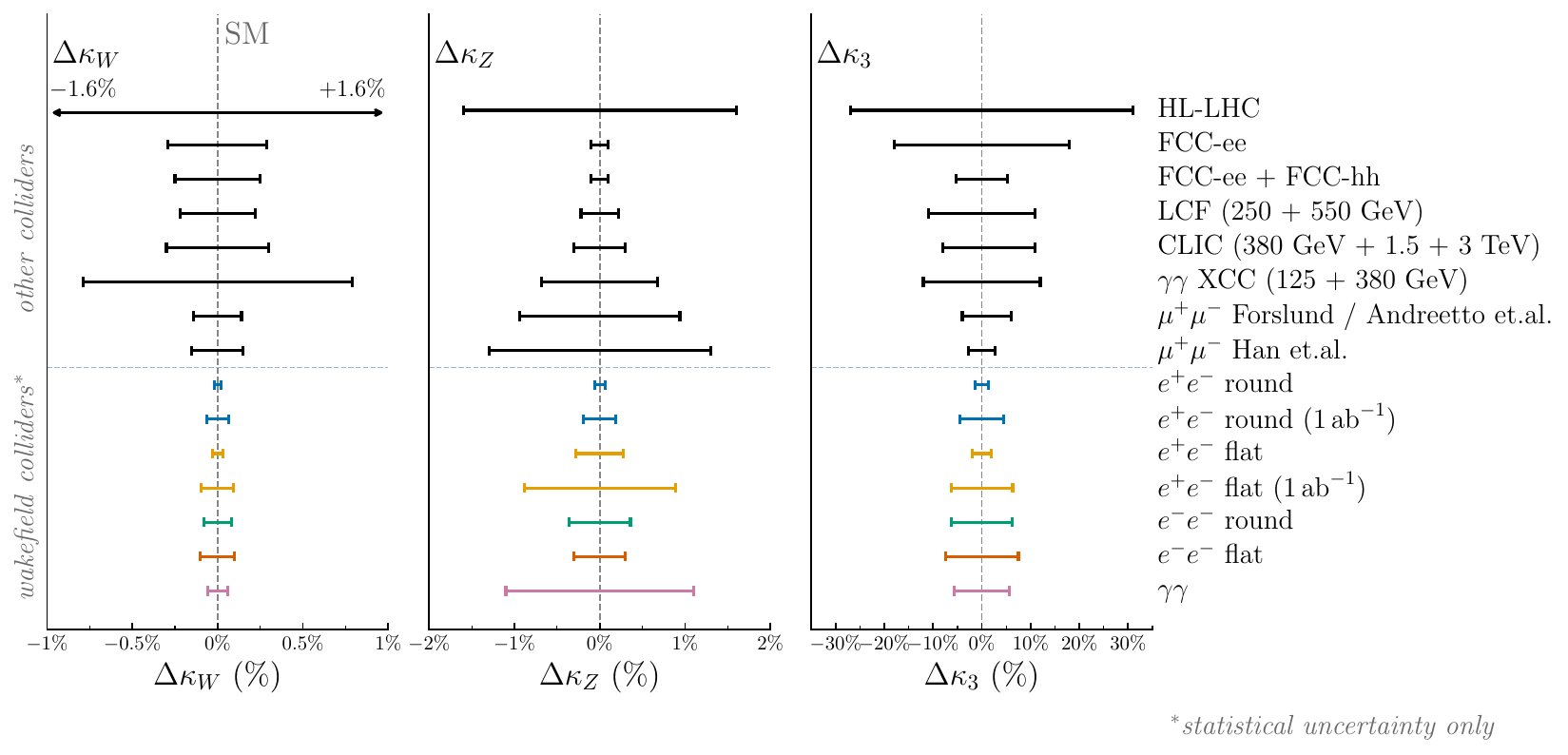}
    \caption{$1\sigma$ sensitivities for $\Delta \kappa_W$, $\Delta \kappa_Z$ and $\Delta\kappa_3$ for the wakefield colliders and a selected set of other future colliders. For $\Delta \kappa_W$ we profiled the likelihood over $\Delta\kappa_Z$ and visa versa; for $\Delta\kappa_3$ we profiled over $\Delta \kappa_{W2}$ and $\Delta \kappa_{Z2}$, setting $\Delta\kappa_{W,Z}=0$.
    We caution against direct comparisons, as the assumptions  between the various colliders differ; see the text for details and references.  }
    \label{fig:intervalplot}
\end{figure*}

In summary, the $e^+e^-$ colliders achieve the best performance owing to their large $W$-boson fusion rates. This conclusion is, however, strongly dependent on the assumed luminosity: if the luminosity of the $e^+e^-$ colliders is reduced by a factor of $\sim10$, reflecting the challenge of positron acceleration, the size of the $e^+e^-$ contours is expected to grow by roughly $\sim \sqrt{10}$. In this case, the advantage of the $e^+e^-$ colliders largely disappears. 
We find no advantage for flat beams over round beams. The reason is that the signal rate is supported over almost the entire center-of-mass range, and for the same geometric luminosity, the total actual luminosity is higher for round beams when integrated over the entire center-of-mass range. 
Furthermore, all vector boson processes involve particles that are either invisible ($\nu_e$) or often outside the forward acceptance of the detector ($W^\pm, e^\pm$).
We therefore find no particular benefit in insisting on beams that are as mono-energetic as possible, as long as the luminosity spectra can be computed or measured to high accuracy.
The $\gamma\gamma$ collider provides a notable improvement over the $e^-e^-$ configurations, as the additional high-energy photons enhance both $\gamma\gamma$- and $e^-\gamma$-initiated signal processes. From the perspective of precision Higgs measurements, such a machine is qualitatively comparable to an $e^+e^-$ wakefield collider operating with roughly one-tenth the luminosity.

To place these results in context, we next compare the wakefield collider projections with those of other future collider proposals to assess their utility as precision instruments.
There are many candidates for such a comparison, whose maturity varies greatly; in \cref{fig:intervalplot} we use the set described below. All sensitivities are at the 1$\sigma$ level.
For the HL-LHC, we use the combined ATLAS/CMS projections assuming $3\,\mathrm{ab}^{-1}$ and
scenario S2~\cite{ATLAS:2025eii}.
For all results with FCC-ee only, and for the FCC-ee$+$FCC-hh combination for $\kappa_{W,Z}$, we show the results from the 2025 FCC Feasibility Study~\cite{FCC:2025lpp}.
For the combined FCC-ee$+$FCC-hh sensitivity to $\kappa_3$, we used the updated FCC-hh studies at 84\,TeV with $30\,\text{ab}^{-1}$, taking their ``scenario 2'' for the systematic uncertainties ~\cite{FCC:2025fub,Stapf:2023ndn}.
For the Linear Collider Facility (LCF) at CERN, we show the full $250+550$\,GeV
program~\cite{LinearCollider:2025lya,LinearColliderVision:2025hlt}.
For CLIC, we use the full three-stage program at 380\,GeV, 1.5\,TeV, and 3\,TeV~\cite{Robson:2020lhl,Roloff:2019crr}.
For the $\gamma\gamma$ XCC, we use  values for the Higgs-pole stage at $\sqrt{s}=125$\,GeV combined with the
380\,GeV stage~\cite{Qureshi:2026ylt,Castelazo:2026iuu}. 
Finally, for the muon collider, we assume a 10\,TeV collider with $10\,\mathrm{ab}^{-1}$ and show two different projections for each coupling. First, we show the recent projections of Forslund et al.~\cite{Forslund:2022xjq} for $\kappa_{W,Z}$, which incorporate HL-LHC information and marginalize over all Higgs couplings, together with the $\kappa_3$ sensitivity from Andreetto et al.~\cite{Andreetto:2026pyu}, which includes a dedicated detector simulation. Second, we also show an older estimate by Han et.al.~\cite{Han:2020pif}. This analysis used a more approximate treatment for the detector and fixed all $\kappa$ parameters to their Standard Model values, except for the shown value, without a combined fit with HL-LHC measurements. 
In this sense, the latter is the closer comparison to our wakefield collider estimates. 
We attribute the differences between our wakefield collider projections and those of Han~et.al.~primarily to the non-trivial luminosity spectra of wakefield colliders.
All wakefield colliders assume a geometric luminosity of $10\,\text{ab}^{-1}$ unless indicated otherwise.

Before drawing conclusions, we want to emphasize that \cref{fig:intervalplot} must be interpreted with care: even leaving aside the great variation in the maturity of the accelerator designs, the intervals are not all obtained under the same assumptions. For instance, the LHC and FCC projections include mature modeling of the detectors and systematic uncertainties, which we have not attempted for the wakefield colliders.
In addition, the systematic uncertainties from the determination of the luminosity spectra still need to be studied.
The statistical treatment itself also differs in some cases since not all analyses perform a simultaneous fit to all Higgs couplings.
Taken together, these caveats imply we cannot conclude from \cref{fig:intervalplot} that a wakefield collider would necessarily be superior to, e.g., FCC-hh or a muon collider.
Nevertheless, while including the aforementioned effects will somewhat degrade the sensitivity, we can conclude from \cref{fig:intervalplot} that wakefield colliders should be capable of precision measurements of the Higgs boson \emph{in the same ballpark} as other collider options. 
In particular, they are likely to further improve upon an LCF or a $\gamma\gamma$ XCC Higgs factory, which would be natural predecessors of a 10 TeV wakefield facility.

In conclusion, we have shown that a 10 TeV wakefield collider would be capable of a competitive Higgs precision program. Flat beams offer no advantage over round beams in this context, and a $\gamma\gamma$ or $e^-e^-$ collider could be an interesting alternative to $e^+e^-$ colliders should positron acceleration remain a bottleneck in the future.
A future 10 TeV wakefield collider would represent a paradigm shift in experimental particle physics, merging the high-precision capabilities of a "factory" with the discovery potential of the energy frontier. As the Standard Model continues to hold under the scrutiny of the LHC, the need for machines that can probe the 1-10 TeV scale with sub-percent precision becomes increasingly urgent.

\begin{acknowledgments}

We gratefully acknowledge the input of collaborators within the LBNL Accelerator Technology \& Applied Physics Division and the LBNL ATLAS group. We are particularly grateful to Stepan Bulanov, Arianna Formenti, R\'emi Lehe, Jens Osterhoff, Carl Schroeder, and Jean-Luc Vay, who generously shared their preliminary findings on accelerator parameters and beam–beam simulations (see \cref{fig:Lumi_Spectra}). We thank Simone Pagan Griso and Angira Rastogi for valuable discussions regarding detector-related assumptions. We likewise thank all of the individuals named above for the ongoing discussions that helped sharpen key concepts and guide the direction of this work. 

We further wish to express our appreciation to Wolfgang Altmanshofer, Tim Barklow, Innes Bigaran, So Chigusa, Daniel Downey, Nathaniel Craig, Spencer Gessner,  Zoltan Ligeti, Pankaj Munbodh, Toby Opferkuch, Michael Peskin, Niki\v{s}a~Ple\v{s}ec, Dean Robinson, Nick Rodd, Christiane Scherb, Inbar Savoray, Linda Xu, and Jure Zupan for their helpful remarks, exchanges, and/or joint work on related topics.

The research of SK is supported by the Office of High Energy Physics of the U.S. Department of Energy under contract DE-AC02-05CH11231. The work of KF is supported by the Miller Institute for Basic Research in Science, University of California Berkeley. The work of  RS was supported by the U.S. Department of Energy under Contract No. DE-SC0012704. Parts of this work were carried out at the Aspen Center for Physics, which receives support from the National Science Foundation under grant PHY-2210452, the Simons Foundation under grant (1161654, Troyer) and the Alfred P. Sloan Foundation under grant (G-2024-22395). Computational resources were provided by the National Energy Research Scientific Computing Center (NERSC), a U.S. Department of Energy Office of Science User Facility, under NERSC award HEP-ERCAP0031191.
\end{acknowledgments}

\newpage 

\bibliographystyle{JHEP}
\bibliography{consolidated_bib.bib}

\appendix
\clearpage 
\onecolumngrid
\setlength{\parskip}{6pt}

\section{Additional Analysis Details\label{app:analysisdetails}}
\vspace{-2mm}
\subsection{Reweighting procedure\label{app:reweight}}
\vspace{-2mm}

All events were generated with \texttt{MadGraph5\_aMC@NLO}~\cite{Alwall:2014hca} on a fixed center of mass energy grid from $250\GeV$ to $10\TeV$ in $50\GeV$ steps. In each case, we compute the number of signal and background events using the reweighting procedure described below. This approach significantly reduces the computational complexity of the analysis, as it allows multiple collider configurations to be studied using the same underlying set of Monte Carlo events. The expected number of events for a particular process is then calculated as follows:

\emph{\textbf{Luminosity Spectrum Weight:}} For any WFC configuration, the \emph{luminosity spectrum weight} for a sample of events at a given center-of-mass energy grid point is defined as:
    \begin{align}
    \label{eq:lumi_weight}
    \mathscr{L}_i = \int_{\sqrt{s}_i - \Delta /2}^{\sqrt{s}_i + \Delta/2}
    \frac{d\mathscr{L}}{d\sqrt{s}}\, d\sqrt{s}\,.
    \end{align}
with $\sqrt{s}_i$ as the center of the bin, $\Delta =50$ GeV as the bin width, and $\frac{d\mathscr{L}}{d\sqrt{s}}$ the luminosity spectra from \cref{fig:Lumi_Spectra}.
Events with $\sqrt{s}=10$ TeV are handled separately to account for the $\delta$-function in the luminosity spectra.  

\emph{\textbf{Decay Weight:}} Using branching ratios $\text{BR}(H \to b\bar{b}) = 0.58$ \cite{LHCHiggsCrossSectionWorkingGroup:2013rie} and $\text{BR}(Z \to b\bar{b}) = 0.15$~\cite{ParticleDataGroup:2024cfk}, we define the \emph{decay weight} as 
\begin{align}
    \mathcal{B} = \text{BR}(H \to b\bar{b})^{N_H} \,\text{BR}(Z \to b\bar{b})^{N_Z}\,.
\end{align}
where $N_{H}$ and $N_Z$ are the numbers of Higgs and Z bosons in a process. For cases where we consider tagging on $W\to jj$, we also include factors of $\text{BR}(W \to jj) \approx 0.67$.

\emph{\textbf{Cross Sections:}} The Standard Model cross sections for all considered signal processes are shown as a function of energy in \cref{fig:Cross_Sections}. As explained in \cref{sec:higgsprecision}, deviations from this cross section can be parameterized using the $\kappa$ framework. For single-Higgs production, the cross-section dependence on $\Delta \kappa_i$ is given in \cref{eq:single_higgs_xsec}. 
For the di-Higgs case, we additionally bin the events in the invariant mass $m_{hh}$ to account for the energy dependence of the operators in \cref{eq:lagrangiankappa}, following the approach in \cite{Han:2020pif}. The di-Higgs production cross section can then be parameterized  as~\cite{Azatov:2015oxa,Bishara:2016kjn,Han:2020pif}
\begin{equation}
\sigma_{ij}^{V}(\Delta\kappa_{V2},\dkthree) = \sigma_{ij}^{V,\text{SM}} \bigl[
  1 + \tilde r_{V2}^{ij} \Delta\kappa_{V2} + \tilde r_{3}^{ij} \dkthree
    + \tilde r_{V2,3}^{ij} \Delta\kappa_{V2}\,\dkthree
    + \tilde r_{V2,V2}^{ij} (\Delta\kappa_{V2})^2
    + \tilde r_{3,3}^{ij} (\dkthree)^2
\bigr],\quad 
\label{eq:sigma_param}
\end{equation}
where $V\in\{W,Z\}$ and the indices $i$ and $j$ denote the center-of-mass energy bin ($\sqrt{s}_i$) and the $m_{hh}$ bin, respectively. 
Alternatively, the number of events after selection cuts and luminosity weighting can be similarly decomposed with its own r-coefficients as
\begin{equation}
N_{ij}^{V}(\Delta\kappa_{V2},\dkthree) = N_{ij}^{V,\text{SM}} \bigl[
  1 + r_{V2}^{ij} \Delta\kappa_{V2} + r_{3}^{ij} \dkthree
    + r_{V2,3}^{ij} \Delta\kappa_{V2}\,\dkthree
    + r_{V2,V2}^{ij} (\Delta\kappa_{V2})^2
    + r_{3,3}^{ij} (\dkthree)^2
\bigr]
\label{eq:N_param}
\end{equation}
As described in \cref{subsec:rcoeffs}, we determine the numerical values of the \(r^{ij}_k\) and \(\tilde r^{ij}_k\) tensors at leading order using MadGraph and our own FeynRules implementation of \cref{eq:lagrangiankappa}. The parametrization in terms of \(\sigma_{ij}^{V}\) is collider independent and can therefore be used to reweight events for arbitrary luminosity spectra. It assumes, however, that variations in the \(\Delta\kappa_i\) do not significantly modify the selection efficiencies relative to their Standard Model values. Conversely, the \(N_{ij}^{V}\) parametrization directly captures changes in the selection efficiencies, but it requires the \( r\)-tensors to be recalculated for each set of luminosity spectra by reweighting a set of MadGraph event samples generated with non-zero $\Delta \kappa_i$ on which the selection cuts have been applied.
We adopt the latter approach for the primary analysis, while verifying that the two methods yield broadly consistent results. Further technical details are provided in \cref{subsec:rcoeffs}. For completeness, we also tabulate both the \(r\) and \(\tilde r\) tensors in that appendix in order to facilitate cross-checks of our analysis and future follow-up studies.

\emph{\textbf{Selection Cuts:}}  
For each event, we apply the selection criteria specified in \cref{tab:cutssingleH} or \cref{tab:cutsdoublehiggs} for the single-Higgs and di-Higgs analyses, respectively. With the exception of the rapidity requirements, all applied cuts are purely transverse. Consequently, they are invariant under boosts along the beamline and remain independent of the relative energy of the incoming particles. This allows for a direct implementation where events are simply assigned a weight of 1 if they pass the selection and 0 otherwise.

The pseudorapidity cuts require more care, given the non-trivial rapidity distribution of the collisions. Specifically, Fig. \ref{fig:Lumi_Spectra} does not show the distribution of the imbalance between the energy of the colliding particles, which determines the boost of the center-of-mass frame relative to the lab frame. 
This boost complicates selection cuts because it may force one or more of the final state particles outside the assumed rapidity acceptance of the detector.
Therefore we must calculate the probability that all final state particles in the event are within the assumed detector acceptance, which depends strongly on the collider and initial state.

We define the conditional rapidity distribution as 
\begin{align}
    p(y|\sqrt{s}_i) = \frac{d\mathscr{L}}{d\sqrt{s}\,dy}\Big|_{\sqrt{s}_i}\Big/ \frac{d\mathscr{L}}{d\sqrt{s}}\Big|_{\sqrt{s}_i}
\end{align}
where $y$ is the rapidity of the center-of-mass frame relative to the lab frame. 
To calculate the probability of all particles in the event being reconstructed, we must integrate this probability distribution over $y$ with boundary conditions 
\begin{equation}
-\eta_{max} < y+\eta_\alpha < \eta_{max}\quad \forall\alpha
\end{equation}
with $\eta_{max}=2.44$ the assumed pseudo-rapidity coverage of the detector and $\eta_\alpha$ the center-of-mass frame pseudo-rapidity of each reconstructed final state particle in a given event.
The probability for a particular event to be fully within the detector acceptance is then 
\begin{align}
w^{(\eta)} =    \int_{-\eta_{max}-\text{min}[\eta_\alpha]}^{\eta_{max}-\text{max}[\eta_\alpha]}
p(y|\sqrt{s}_i) \, dy.
\end{align}
The weight $w^{(\eta)}$ is then applied to each event independently in both the calculation of the Standard Model cross section and of the r-coefficients.

\emph{\textbf{The total event yield:}} The total event yield is then given by 
\begin{equation} 
    N_{(j)} (\Delta \kappa) = \mathcal{B} \sum_i \sigma_{i(j)}(\Delta \kappa) \mathscr{L}_i 
\end{equation}
where the index $i$ again sums over the center-of-mass energy bin, and the index $j$ only applies for the di-Higgs analysis, where it denotes the $m_{hh}$ bin. 

\vspace{-4mm}
\subsection{Forward Detector Coverage Dependence\label{app:coverage}}
\vspace{-1mm}

\begin{figure*}[t!]
    \centering
    \includegraphics[width=0.8\linewidth]{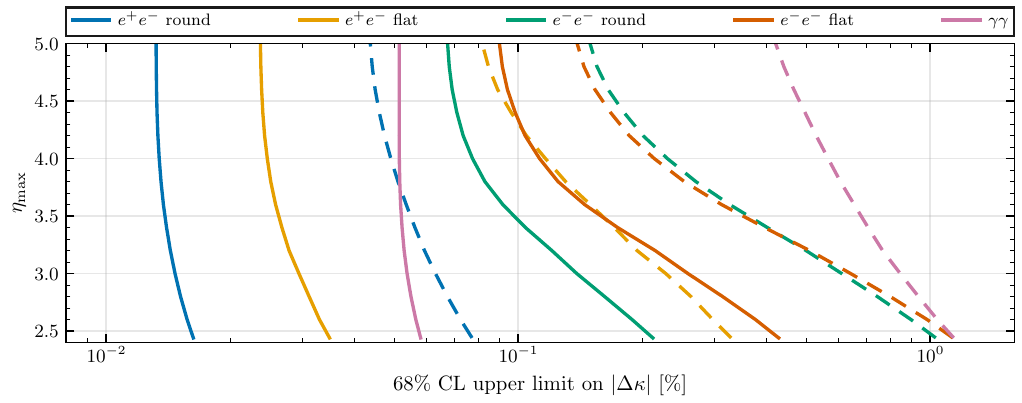}
    \caption{Single-Higgs coupling sensitivity (68\% C.L.) as a function of detector rapidity coverage $\eta_{\max}$, for all five \WFC\ configurations. Solid lines show $\dkW$ and dashed lines show $\dkZ$. The $e^-e^-$ configurations benefit most from extended forward coverage, which breaks the WBF/ZBF degeneracy.}
    \label{fig:sensitivity_vs_eta_max}
\end{figure*}

To quantify the physics gain for the single Higgs analysis from extending forward detector coverage we consider modifying $\eta_{\rm det} \in [2.44,\,5.0]$, holding the other selection cuts of \cref{tab:cutssingleH} fixed. Figure~\ref{fig:sensitivity_vs_eta_max} shows resulting $68\%$~CL sensitivities. The largest gains are for $e^-e^-$ configurations, where the baseline acceptance leaves WBF and ZBF nearly degenerate. The $\gamma\gamma$ collider behaves similarly. The $e^+e^-$ configurations gain a further $25\text{--}35\%$ from additional $e^\pm$ tagging efficiency. 

The figure assumes that lepton and jet tagging at $|\eta|<\eta_{\max}$ is achievable, however, coherent and incoherent pair production and $\gamma\gamma\to\,$ hadrons also populate forward region and can complicate detection there.  The pseudo-rapidity range over which this can be achieved is a question of the design of the detector, the constraints from the beam delivery system and the degree to which beam-induced backgrounds can be rejected.

\vspace{-4mm}
\subsection{Quantifying the Effect of Tagging\label{app:tag}}

\begin{figure}[t!]
    \centering

    \includegraphics[width=0.75\linewidth]{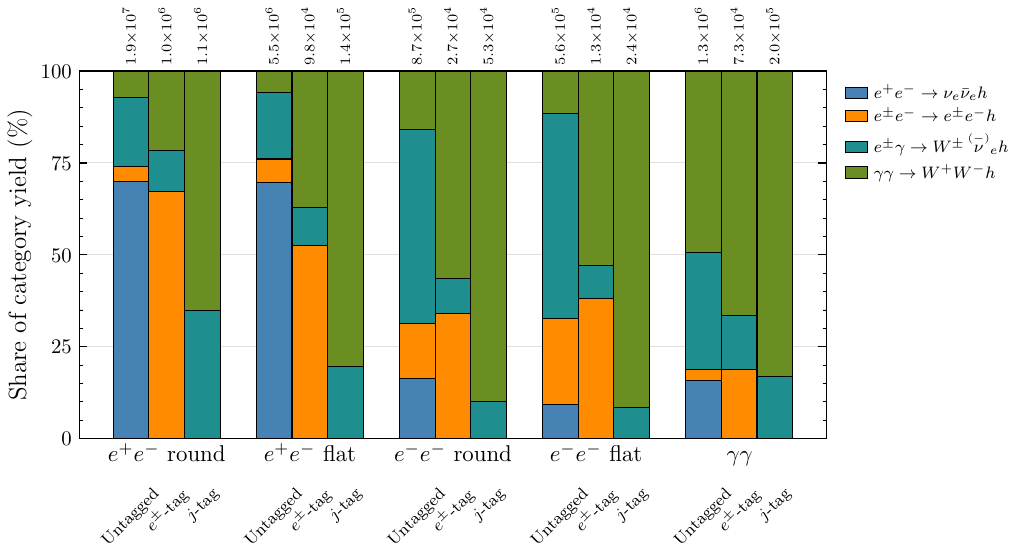}
    
    \vspace{1ex}
    \includegraphics[width=0.8\linewidth]{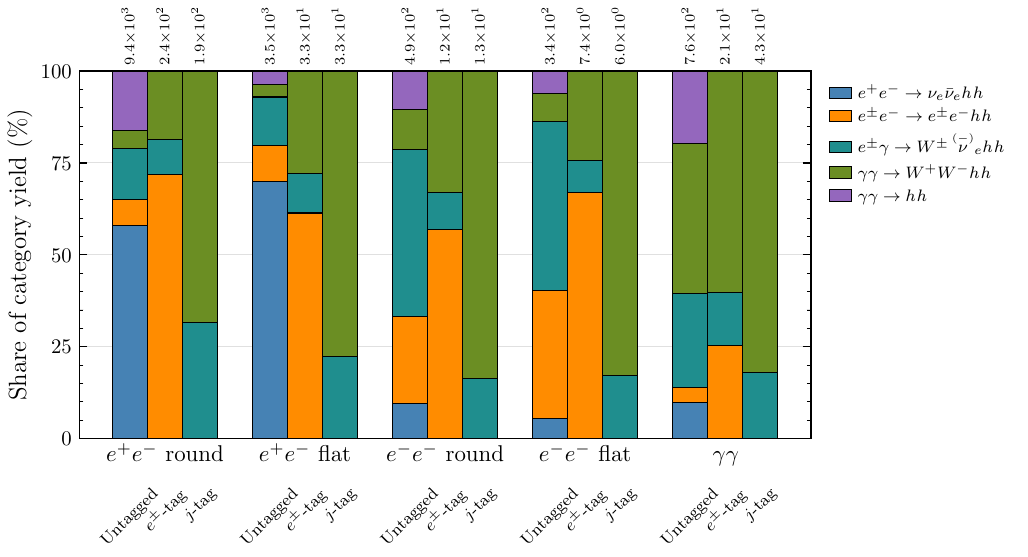}

    \caption{Event yields at an integrated luminosity of $10\,\mathrm{ab}^{-1}$ are shown for the WFC configurations, separated into untagged, $e^\pm$-tagged, and $j$-tagged categories. $h \to b\bar b$ branching ratios are included. The top panel corresponds to single-Higgs production, while the bottom panel corresponds to double-Higgs production. In both panels, the total event count is indicated at the top of each plot.}
    \label{fig:single_and_doublehiggs_tagging_stacked}
\end{figure}

To quantify the impact of tagging on the relative importance of various single and double-Higgs production channels, we analyze both the integrated and differential event yields. We do this by evaluating the breakdown of the total event yield by process alongside the differential yield as a function of the effective center-of-mass energy, $\sqrt{s}$.

The process-by-process breakdown of the total event yield is shown in \cref{fig:single_and_doublehiggs_tagging_stacked}, with the top and bottom rows corresponding to single- and double-Higgs production, respectively. For each collider configuration, the three adjacent bars correspond, from left to right, to the untagged, electron-tagged, and jet-tagged selections. Comparing these distributions reveals substantial differences in the process composition of each tagging category. In the untagged category, $e^+e^- \to \nu_e\overline{\nu}_e h$ dominates for $e^+e^-$ colliders, while $e^-\gamma \to W^{\pm}\nu_e h$ dominates for $e^-e^-$ colliders and $\gamma\gamma \to W^+W^-h$ provides the largest contribution for $\gamma\gamma$ colliders. In the electron-tagged category, $e^+e^- \to e^+e^-h$ provides the dominant single-Higgs contribution for $e^+e^-$ beams, while $e^{\pm}e^- \to e^{\pm}e^-hh$ dominates double-Higgs production for both $e^+e^-$ and $e^-e^-$ beams. By contrast, for single-Higgs production with $e^-e^-$ beams, as well as for $\gamma\gamma$ colliders, $\gamma\gamma \to W^+W^-h$ provides the largest contribution after imposing an electron tag. In the jet-tagged category, $\gamma\gamma \to W^+W^-h$ is consistently the dominant contribution across all collider configurations.

\begin{figure}[t]
    \centering
    \includegraphics[width=\linewidth]{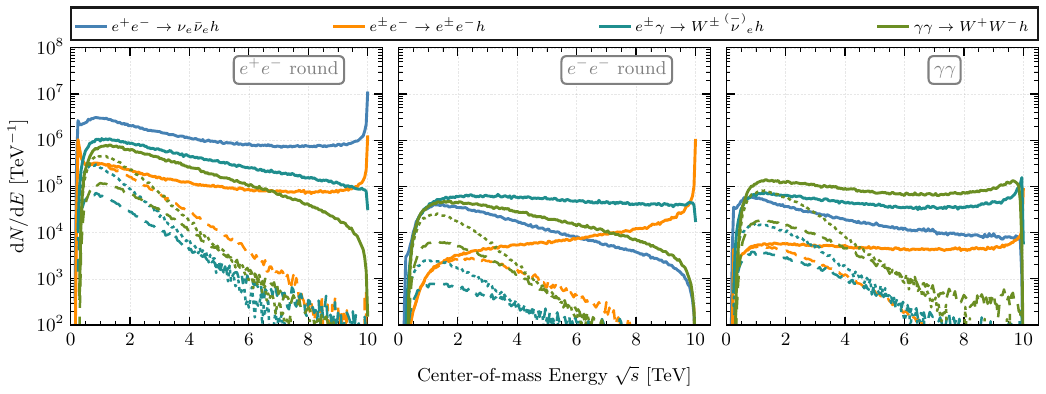}
    \vspace{1ex}
    \includegraphics[width=\linewidth]{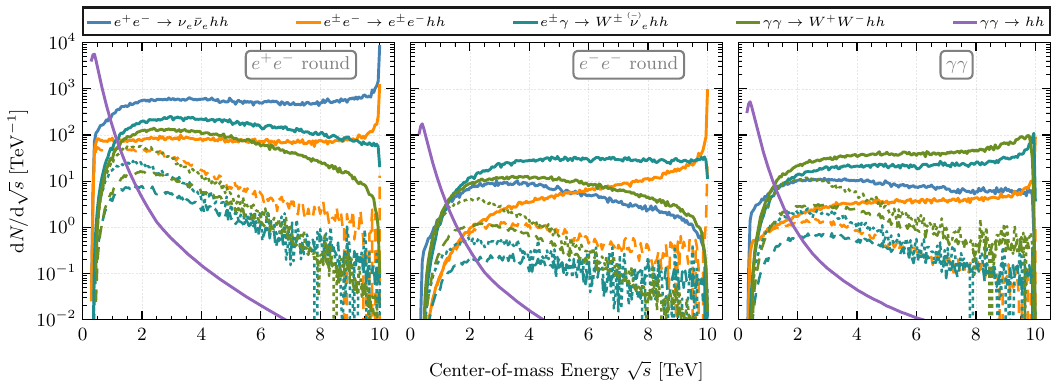}
    
    \caption{Differential event yields for single-Higgs (top) and di-Higgs (bottom) production, including the branching ratio $\mathrm{BR}(H\to b\bar b)$, as a function of the center-of-mass energy for three WFC configurations: $e^+e^-$ round, $e^-e^-$ round, and $\gamma\gamma$, assuming an integrated luminosity of $10\,\mathrm{ab}^{-1}$. Colors indicate the production channel, with solid curves for the total number of events, dashed curves for $e$-tagged events, and dotted curves for jet-tagged events.}
    \label{fig:sh_and_dh_events_vs_energy_stacked}
\end{figure}

The corresponding differential event yields as a function of $\sqrt{s}$, including the cuts from \cref{tab:cutssingleH} and \cref{tab:cutsdoublehiggs}, are shown are displayed in \cref{fig:sh_and_dh_events_vs_energy_stacked}. Here, the production channels are distinguished by color, while the line styles encode the tagging requirement, using solid curves for the total number of events, dashed curves for $e$-tagged events, and dotted curves for jet-tagged events. As expected, the $\gamma\gamma \rightarrow hh$ cross section peaks sharply near threshold and falls rapidly as $\sqrt{s}$ increases. Conversely, the $e^{\pm}e^- \rightarrow e^{\pm}e^-h$ channel exhibits a pronounced peak at $\sqrt{s}\sim 10~\mathrm{TeV}$ for both $e^+e^-$ and $e^-e^-$ beam configurations, and the $e^+e^- \rightarrow \nu_e\overline{\nu}_eh$ channel features a stark spike at $\sqrt{s}\sim 10~\mathrm{TeV}$ in the $e^+e^-$ configuration, reflecting the corresponding $\delta$-function luminosity spike for these initial states.
The $e$ and jet-tagging are most effective at low $\sqrt{s}$, which means round beams benefit the most.

\vspace{-3mm}
\subsection{r-Coefficients}
\label{subsec:rcoeffs}
\vspace{-1mm}

In this subsection, we first describe the calculation of the $\tilde r$-coefficients, which are evaluated independently in each $\sqrt{s}$ and $m_{hh}$ bin and can therefore be convolved with arbitrary luminosity spectra. We also tabulate the resulting coefficients for each process as functions of $\sqrt{s}$ and $m_{hh}$. We then calculate the corresponding $r$-coefficients for the primary analysis, incorporating the collider-specific luminosity spectra and selection efficiencies, and give their numerical values for each collider considered in this paper.

The procedure used to calculate the $\tilde r$-coefficients is as follows. We evaluate the total cross section on a grid of nine $(\Delta\kappa_{V2},\Delta\kappa_3)$ points, generating $10^5$ events in each relevant $m_{hh}$ interval. The upper boundary of each interval is imposed through a custom generator-level cut in MadGraph. We take $\kappa_V=1$, with $\Delta\kappa_{V2}=-1.0,\,0,\,0.5$ and $\Delta\kappa_3=-0.5,\,0,\,1.0$. The relatively large coupling variations are chosen to ensure that the deviation from the SM is large compared to the statistical uncertainties in the MadGraph samples.

We then solve for the $\tilde r$-coefficients using \cref{eq:sigma_param}, treating each signal process independently. We first determine the coefficients associated with variations in a single $\Delta\kappa_i$, using the two corresponding BSM points together with the SM point, while fixing the remaining coupling deviations to their Standard Model values. We then extract the mixed term from the remaining four parameter points. The resulting four estimates of the mixed $\tilde r$-coefficient are combined using inverse-variance weighting, where the uncertainty associated with each of these four points is
\begin{equation}
    \sigma^{\tilde r}
    =
    \frac{\delta\sigma}
    {\sigma_{\rm SM}
    \left|\Delta\kappa_{V2}\Delta\kappa_{3}\right|},
\end{equation}
and $\delta \sigma$ is the MadGraph statistical uncertainty on the cross section. The uncertainties on the resulting $\tilde r$-coefficients are obtained by propagating the MadGraph cross-section uncertainties from all nine parameter points.

\begin{figure}[p]
  \centering
  \vspace*{-9mm}
  \hspace*{-1.0cm}
  \includegraphics[
    width=1.05\linewidth
  ]{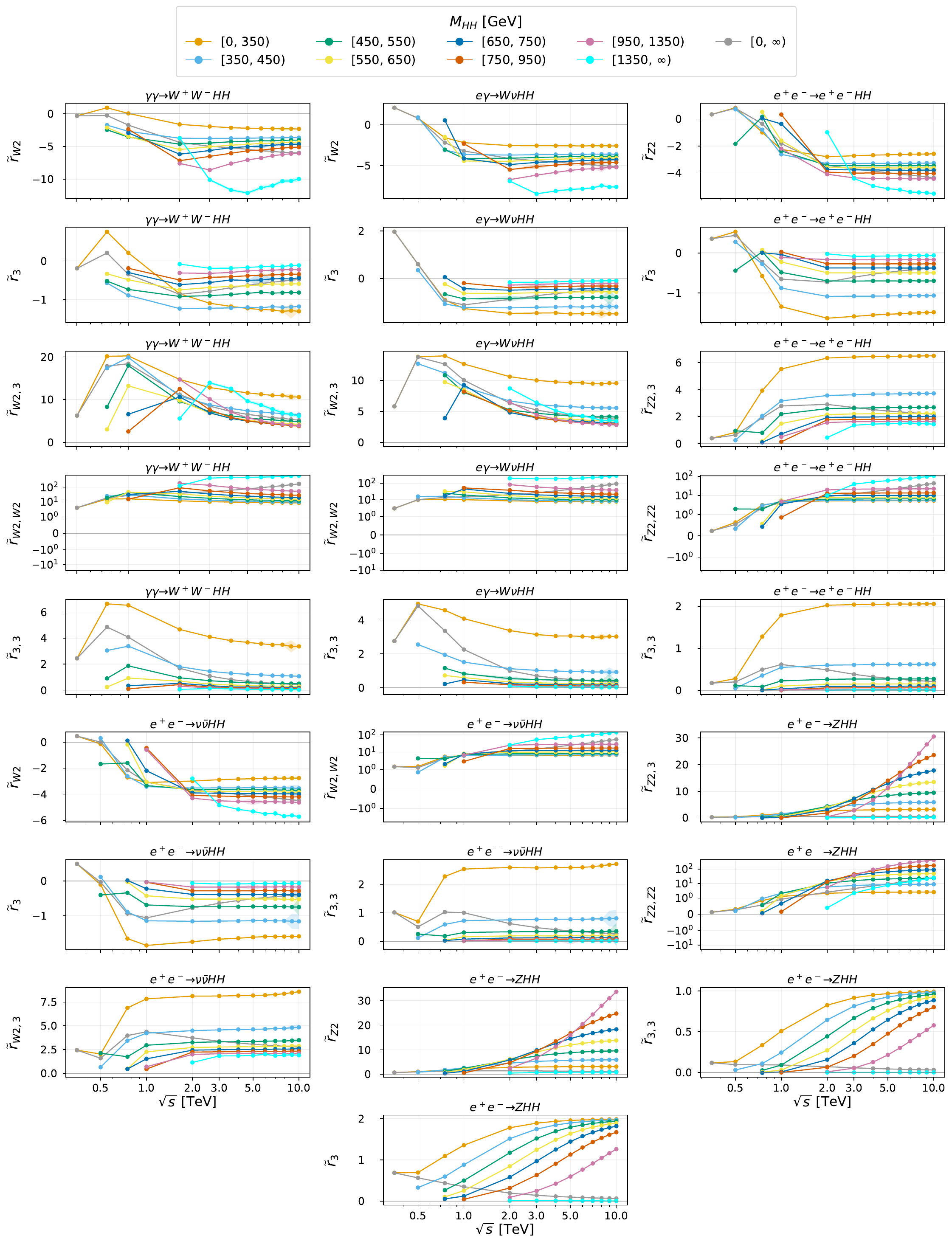}

  \begin{tikzpicture}[remember picture,overlay]
    \node[
      anchor=north east,
      inner sep=0pt
    ] at ([xshift=-19mm,yshift=-9mm]current page.north east) {
      \normalfont\pageref*{fig:r_coeffs_with_errs}
    };
  \end{tikzpicture}

  \vspace*{-6mm}
  \caption{The \(\tilde r\) coefficients from \cref{eq:sigma_param} and their uncertainties are shown as functions of \(\sqrt{s}\) for each process and coefficient, both inclusively and in each \(m_{HH}\) bin. The MadGraph statistical uncertainties are typically too small to be visible and remain well controlled.}
  \label{fig:r_coeffs_with_errs}
\end{figure}
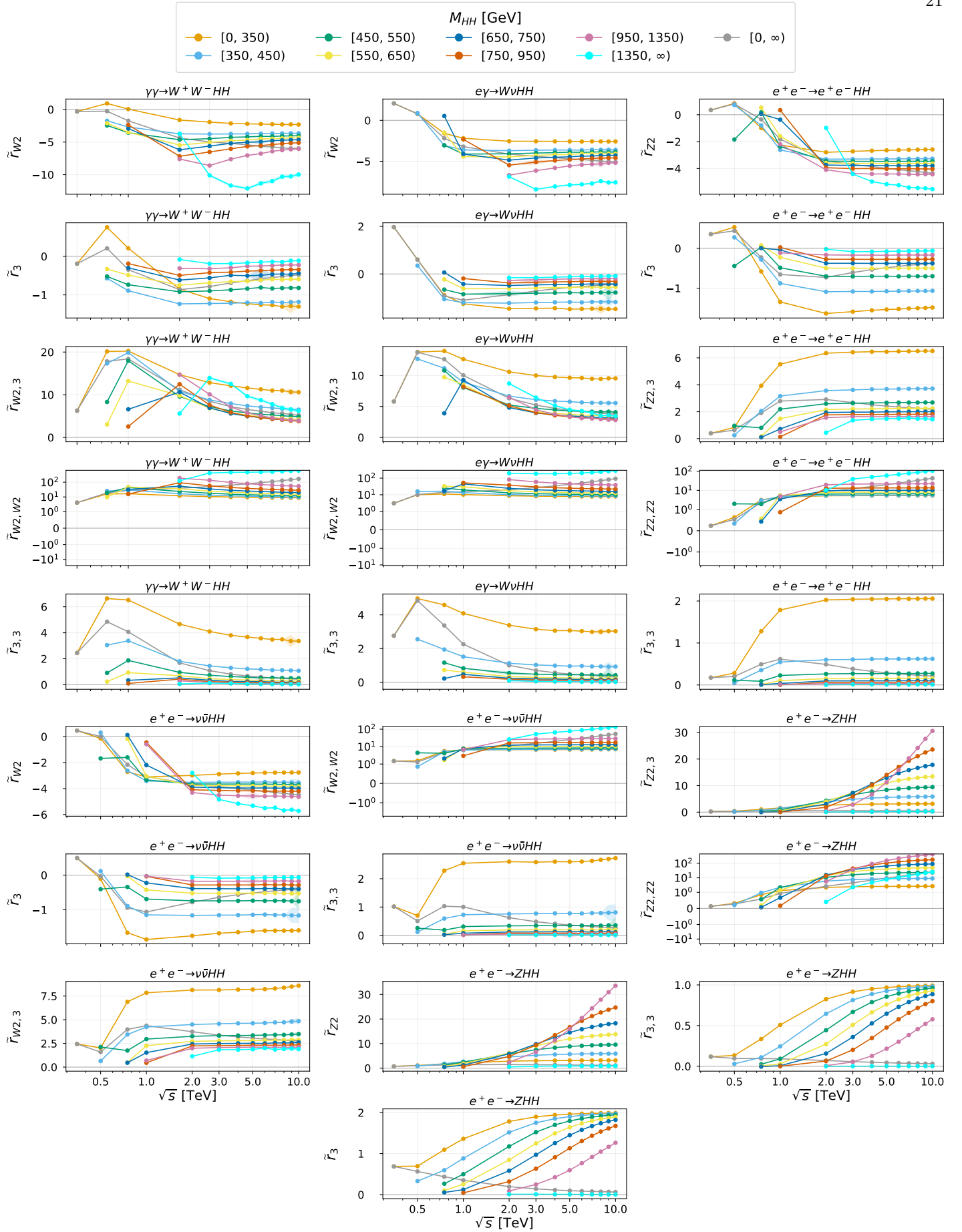

The resulting $\tilde r$-coefficients are shown as functions of $\sqrt{s}$ for each process, with a separate colored curve for each $m_{hh}$ bin, in \cref{fig:r_coeffs_with_errs}. The points marked by dots are evaluated directly in MadGraph, and linear interpolation is used between adjacent values of $\sqrt{s}$. The uncertainties on the $\tilde{r}$-coefficients are indicated by shaded bands.  In most cases, the uncertainty bands are too small to be visible, demonstrating that the statistical uncertainties on the $\tilde r$-coefficients are negligible. The few visible exceptions generally occur in regions of phase space that are difficult to sample, where the custom cut does not modify MadGraph's phase-space sampler and therefore leads to somewhat larger uncertainties. Overall, the uncertainties on the $\tilde r$-coefficients remain small and well controlled. For completeness, the coefficients and their uncertainties are also given in Tables \ref{tab:r_coeffs_aa_wwhh}--\ref{tab:r_coeffs_ee_eehh_samesign}. As discussed in \cref{app:analysisdetails}, these values could be used to 
used to reweight SM Monte Carlo samples for different collider and detector assumptions..

The $r$-coefficients in \cref{eq:N_param} are calculated using a similar, but slightly simpler, procedure. In this case, a single point with both $\Delta\kappa_{V2}$ and $\Delta\kappa_3$ nonzero is used to extract $r_{V2,3}$. Event yields are obtained by summing over $\sqrt{s}$, weighted by the luminosity spectra and the selection efficiencies defined in \cref{app:analysisdetails}. Here, we neglect variations within each energy bin and evaluate the cross section at the bin center, an approximation that is adequate for the chosen bin width \(\Delta E\). Samples are generated either independently at each point in the $\kappa_i$ parameter space or, where adequate, using MadGraph's built-in reweighting procedure to reduce the computational cost. For each collider and $m_{hh}$ bin, the $r$-coefficients are obtained by summing the contributions from all relevant processes. The six points used to extract the coefficients are
\begin{equation}
    (\Delta\kappa_{V2},\Delta\kappa_3)
    ={}(0,0),\,(-0.5,0),\,(0.5,0),(0,2.0),\,(0,-2.0),\,(0.5,2.0),
\end{equation}
all with $\kappa_V=1$. We use this set of coefficients in the full analysis because, once the luminosity spectrum is specified, they provide the most precise parametrization due to incorporating changes in the selection efficiencies induced by variations in the $\Delta \kappa_i$.

\begin{figure}[p]
    \centering
    \includegraphics[width=.8\linewidth]{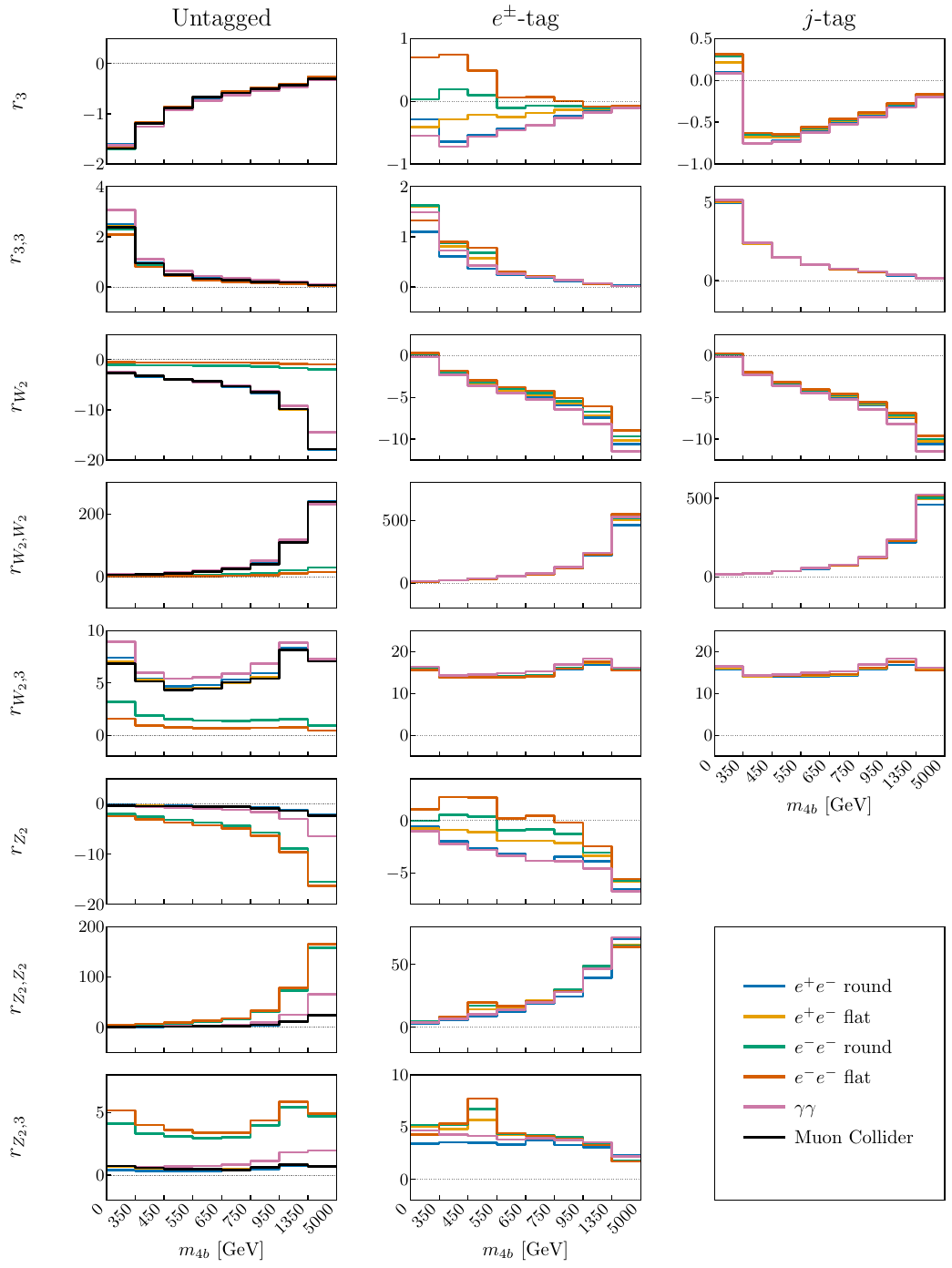}
    \caption{The \(r\)-coefficients defined in \cref{eq:N_param} are shown as functions of \(m_{4b}\) for each tagging category. The contributions from \(\sqrt{s}\) are summed using the corresponding luminosity profiles, with color indicating the collider configuration.}
    \label{fig:rcoeffs_including_lumi}
\end{figure}

The resulting $r$-coefficients are shown in \cref{fig:rcoeffs_including_lumi} as functions of $m_{4b}$. Each tagging category is displayed separately, and color denotes the collider configuration. In many cases, the coefficients are similar across the different colliders. The main exceptions are coefficients involving $V_2$ in the untagged category, for which the $e^-e^-$ colliders differ from the $e^+e^-$ and $\gamma\gamma$ colliders because of the lower rate for WBF, and $r_3$, $r_{Z2}$, and $r_{Z2,3}$ in the $e^{\pm}$-tagged category, for which substantial differences are observed across the collider configurations. For reproducibility, these coefficients are also provided in \cref{tab:r_coeffs_binned_allinone}. 

\newcommand{\compacttimes}{\mathord{\scalebox{0.75}{$\times$}}}
\newcommand{\compactminus}{\scalebox{0.64}[0.82]{$\scriptstyle-$}}
\newcommand{\compactpm}{\mathord{\scalebox{0.78}{$\pm$}}}

\newcommand{\pair}[4]{%
  \ifnum\numexpr#2+3\relax=0
    $#1\compactpm\abserror{#3}{#4}$%
  \else
    \ifnum\numexpr#2+3\relax=1
      $\fpeval{#1*10}.0\compactpm\abserror{#3}{#4}$%
    \else
      \ifnum\numexpr#2+3\relax=2
        $\fpeval{#1*100}\compactpm\abserror{#3}{#4}$%
      \else
        \ifnum\numexpr#2+3\relax=-1
          $\expandafter\minusone#1\relax\compactpm\abserror{#3}{#4}$%
        \else
          \ifnum\numexpr#2+3\relax=-2
            $\expandafter\minustwo#1\relax\compactpm\abserror{#3}{#4}$%
          \else
            \ifnum\numexpr#2+3\relax=-3
              $\expandafter\minusthree#1\relax\compactpm\abserror{#3}{#4}$%
            \else
              $(#1\compactpm\fpeval{#3*10^(#4-#2)})\compacttimes10^{\number\numexpr#2+3\relax}$%
            \fi
          \fi
        \fi
      \fi
    \fi
  \fi
}
\def\minusone#1.#2\relax{0.#1#2}
\def\minustwo#1.#2\relax{0.0#1#2}
\def\minusthree#1.#2\relax{0.00#1#2}
\newcommand{\abserror}[2]{%
  \ifnum\numexpr#2+3\relax<-3
    #1\compacttimes10^{\compactminus\expandafter\errorexp#2\relax}%
  \else
    \fpeval{#1*10^(#2+3)}%
  \fi
}
\def\errorexp-#1\relax{\number\numexpr#1-3\relax}

\begin{table*}[p]
\vspace*{-6mm}
\makebox[\textwidth][c]{%
  \hspace*{-0.9cm}%
  \begin{minipage}{\textwidth}
\centering
\caption{$\tilde{r}$-coefficients, SM cross sections, and their errors for the inclusive $\gamma\gamma \to W^+W^-hh$ cross section.}
\setlength{\belowcaptionskip}{0pt}
\label{tab:r_coeffs_aa_wwhh}
\vspace{-2mm}
\scriptsize
\setlength{\tabcolsep}{1pt}
\begin{tabular}{cc|ccccccccc}
\toprule
\textbf{Coeff.} & \shortstack{$\boldsymbol{\sqrt{s}}$\\[-1pt] [TeV]} & \textbf{Total} & \textbf{0--350} & \textbf{350--450} & \textbf{450--550} & \textbf{550--650} & \textbf{650--750} & \textbf{750--950} & \textbf{950--1350} & \textbf{1350}\,{\raisebox{0.2ex}{\tiny$\boldsymbol{+}$}} \\
\midrule
\multirow{12}{*}{$\sigma_{\mathrm{SM}}\,[\mathrm{ab}]$} & 0.50 & \pair{2.4}{-4}{1.5}{-7} & \pair{2.4}{-4}{1.5}{-7} & --- & --- & --- & --- & --- & --- & --- \\
~ & 0.75 & \pair{5.3}{-3}{4.8}{-6} & \pair{3.1}{-3}{3.2}{-6} & \pair{1.6}{-3}{1.6}{-6} & \pair{5.9}{-4}{7.0}{-7} & \pair{2.7}{-5}{3.3}{-8} & --- & --- & --- & --- \\
~ & 1.00 & \pair{2.0}{-2}{2.0}{-5} & \pair{8.4}{-3}{9.1}{-6} & \pair{5.9}{-3}{6.9}{-6} & \pair{3.1}{-3}{4.0}{-6} & \pair{1.6}{-3}{1.8}{-6} & \pair{8.1}{-4}{1.2}{-6} & \pair{1.7}{-4}{2.3}{-7} & --- & --- \\
~ & 2.00 & \pair{2.1}{-1}{2.2}{-4} & \pair{4.1}{-2}{3.9}{-5} & \pair{5.0}{-2}{5.4}{-5} & \pair{4.0}{-2}{4.4}{-5} & \pair{2.7}{-2}{2.4}{-5} & \pair{1.8}{-2}{2.1}{-5} & \pair{1.8}{-2}{2.4}{-5} & \pair{1.1}{-2}{1.0}{-5} & \pair{2.9}{-3}{4.1}{-6} \\
~ & 3.00 & \pair{5.4}{-1}{6.5}{-4} & \pair{7.1}{-2}{8.1}{-5} & \pair{9.9}{-2}{1.3}{-4} & \pair{9.0}{-2}{1.0}{-4} & \pair{7.2}{-2}{8.2}{-5} & \pair{5.4}{-2}{4.9}{-5} & \pair{6.9}{-2}{7.7}{-5} & \pair{5.3}{-2}{5.4}{-5} & \pair{2.1}{-2}{2.3}{-5} \\
~ & 4.00 & \pair{9.2}{-1}{1.1}{-3} & \pair{9.6}{-2}{1.2}{-4} & \pair{1.4}{-1}{2.0}{-4} & \pair{1.4}{-1}{1.5}{-4} & \pair{1.2}{-1}{1.5}{-4} & \pair{9.4}{-2}{1.1}{-4} & \pair{1.3}{-1}{1.4}{-4} & \pair{1.2}{-1}{1.3}{-4} & \pair{6.6}{-2}{7.3}{-5} \\
~ & 5.00 & \pair{1.3}{0}{1.5}{-3} & \pair{1.2}{-1}{1.1}{-4} & \pair{1.8}{-1}{2.2}{-4} & \pair{1.8}{-1}{2.2}{-4} & \pair{1.6}{-1}{1.9}{-4} & \pair{1.3}{-1}{1.4}{-4} & \pair{1.9}{-1}{2.1}{-4} & \pair{2.0}{-1}{2.4}{-4} & \pair{1.4}{-1}{1.9}{-4} \\
\bottomrule
\end{tabular}
\end{minipage}%
}
\end{table*}

\begin{table*}[p]
\vspace*{-6mm}
\makebox[\textwidth][c]{%
  \hspace*{-2.5cm}%
  \begin{minipage}{\textwidth}
\centering
\caption{$\tilde{r}$-coefficients, SM cross sections, and their errors for the inclusive $e\gamma \to W\nu hh$ cross section.}
\setlength{\belowcaptionskip}{0pt}
\label{tab:r_coeffs_ea_wnuhh}
\vspace{-2mm}
\scriptsize
\setlength{\tabcolsep}{1pt}
\begin{tabular}{cc|ccccccccc}
\toprule
\textbf{Coeff.} & \shortstack{$\boldsymbol{\sqrt{s}}$\\[-1pt] [TeV]} & \textbf{Total} & \textbf{0--350} & \textbf{350--450} & \textbf{450--550} & \textbf{550--650} & \textbf{650--750} & \textbf{750--950} & \textbf{950--1350} & \textbf{1350}\,{\raisebox{0.2ex}{\tiny$\boldsymbol{+}$}} \\
\midrule
\multirow{13}{*}{$\sigma_{\mathrm{SM}}\,[\mathrm{ab}]$} & 0.35 & \pair{1.8}{-7}{1.4}{-10} & \pair{1.8}{-7}{1.4}{-10} & --- & --- & --- & --- & --- & --- & --- \\
~ & 0.50 & \pair{4.3}{-4}{3.7}{-7} & \pair{4.1}{-4}{4.0}{-7} & \pair{1.9}{-5}{1.9}{-8} & --- & --- & --- & --- & --- & --- \\
~ & 0.75 & \pair{6.9}{-3}{8.6}{-6} & \pair{3.9}{-3}{3.3}{-6} & \pair{2.3}{-3}{1.8}{-6} & \pair{5.6}{-4}{7.1}{-7} & \pair{5.8}{-5}{6.5}{-8} & \pair{1.1}{-7}{7.5}{-11} & --- & --- & --- \\
~ & 1.00 & \pair{2.6}{-2}{3.5}{-5} & \pair{9.8}{-3}{7.6}{-6} & \pair{9.1}{-3}{1.3}{-5} & \pair{4.7}{-3}{6.1}{-6} & \pair{1.8}{-3}{2.2}{-6} & \pair{4.9}{-4}{4.8}{-7} & \pair{9.8}{-5}{1.0}{-7} & --- & --- \\
~ & 2.00 & \pair{2.1}{-1}{2.4}{-4} & \pair{3.5}{-2}{3.9}{-5} & \pair{4.9}{-2}{5.9}{-5} & \pair{4.2}{-2}{4.7}{-5} & \pair{3.1}{-2}{1.8}{-5} & \pair{2.1}{-2}{1.6}{-5} & \pair{2.2}{-2}{1.8}{-5} & \pair{1.1}{-2}{1.5}{-5} & \pair{8.9}{-4}{1.2}{-6} \\
~ & 3.00 & \pair{4.9}{-1}{5.3}{-4} & \pair{5.6}{-2}{6.5}{-5} & \pair{8.4}{-2}{8.3}{-5} & \pair{8.0}{-2}{8.6}{-5} & \pair{6.6}{-2}{7.4}{-5} & \pair{5.1}{-2}{5.5}{-5} & \pair{6.8}{-2}{4.8}{-3} & \pair{5.7}{-2}{3.3}{-5} & \pair{1.9}{-2}{2.2}{-5} \\
~ & 4.00 & \pair{7.7}{-1}{9.3}{-4} & \pair{7.1}{-2}{8.3}{-5} & \pair{1.1}{-1}{1.3}{-4} & \pair{1.1}{-1}{1.4}{-4} & \pair{9.6}{-2}{1.2}{-4} & \pair{7.9}{-2}{1.0}{-4} & \pair{1.1}{-1}{1.2}{-4} & \pair{1.2}{-1}{9.2}{-5} & \pair{6.7}{-2}{7.8}{-5} \\
\bottomrule
\end{tabular}
\end{minipage}%
}
\end{table*}

\begin{table*}[p]
\vspace*{-6mm}
\makebox[\textwidth][c]{%
  \hspace*{-1.0cm}%
  \begin{minipage}{\textwidth}
\centering
\caption{$\tilde{r}$-coefficients, SM cross sections, and their errors for the inclusive $e^+e^- \to e^+e^-hh$ cross section.}
\setlength{\belowcaptionskip}{0pt}
\label{tab:r_coeffs_ee_eehh}
\vspace{-2mm}
\scriptsize
\setlength{\tabcolsep}{1pt}
\begin{tabular}{cc|ccccccccc}
\toprule
\textbf{Coeff.} & \shortstack{$\boldsymbol{\sqrt{s}}$\\[-1pt] [TeV]} & \textbf{Total} & \textbf{0--350} & \textbf{350--450} & \textbf{450--550} & \textbf{550--650} & \textbf{650--750} & \textbf{750--950} & \textbf{950--1350} & \textbf{1350}\,{\raisebox{0.2ex}{\tiny$\boldsymbol{+}$}} \\
\midrule
\multirow{13}{*}{$\sigma_{\mathrm{SM}}\,[\mathrm{ab}]$} & 0.35 & \pair{1.7}{-4}{1.6}{-7} & \pair{1.7}{-4}{1.6}{-7} & --- & --- & --- & --- & --- & --- & --- \\
~ & 0.50 & \pair{5.7}{-3}{5.7}{-6} & \pair{3.7}{-3}{4.1}{-6} & \pair{2.0}{-3}{2.1}{-6} & \pair{1.1}{-6}{1.2}{-9} & --- & --- & --- & --- & --- \\
~ & 0.75 & \pair{8.5}{-3}{7.2}{-6} & \pair{2.5}{-3}{1.8}{-6} & \pair{2.4}{-3}{2.1}{-6} & \pair{1.9}{-3}{2.5}{-6} & \pair{1.8}{-3}{1.9}{-6} & \pair{8.7}{-5}{1.2}{-7} & --- & --- & --- \\
~ & 1.00 & \pair{1.4}{-2}{8.2}{-6} & \pair{3.3}{-3}{3.0}{-6} & \pair{3.6}{-3}{3.3}{-6} & \pair{2.5}{-3}{1.9}{-6} & \pair{1.5}{-3}{1.6}{-6} & \pair{1.1}{-3}{1.4}{-6} & \pair{1.7}{-3}{2.0}{-6} & \pair{2.3}{-7}{2.5}{-10} & --- \\
~ & 2.00 & \pair{5.6}{-2}{6.2}{-5} & \pair{7.6}{-3}{1.0}{-5} & \pair{1.1}{-2}{1.4}{-5} & \pair{1.0}{-2}{1.2}{-5} & \pair{8.0}{-3}{1.0}{-5} & \pair{5.8}{-3}{7.3}{-6} & \pair{7.0}{-3}{9.5}{-6} & \pair{4.6}{-3}{5.8}{-6} & \pair{1.8}{-3}{1.6}{-6} \\
~ & 3.00 & \pair{1.1}{-1}{1.6}{-4} & \pair{1.1}{-2}{1.5}{-5} & \pair{1.7}{-2}{2.2}{-5} & \pair{1.7}{-2}{2.0}{-5} & \pair{1.4}{-2}{1.8}{-5} & \pair{1.2}{-2}{1.5}{-5} & \pair{1.6}{-2}{2.0}{-5} & \pair{1.6}{-2}{2.1}{-5} & \pair{7.9}{-3}{1.0}{-5} \\
~ & 4.00 & \pair{1.7}{-1}{2.4}{-4} & \pair{1.3}{-2}{1.6}{-5} & \pair{2.2}{-2}{2.9}{-5} & \pair{2.2}{-2}{2.6}{-5} & \pair{2.0}{-2}{2.4}{-5} & \pair{1.7}{-2}{2.2}{-5} & \pair{2.5}{-2}{2.8}{-5} & \pair{2.8}{-2}{3.0}{-5} & \pair{2.1}{-2}{2.6}{-5} \\
\bottomrule
\end{tabular}
\end{minipage}%
}
\end{table*}

\begin{table*}[p]
\vspace*{-6mm}
\makebox[\textwidth][c]{%
  \hspace*{-1.4cm}%
  \begin{minipage}{\textwidth}
\centering
\caption{$\tilde{r}$-coefficients, SM cross sections, and their errors for the inclusive $e^+e^- \to \nu\bar{\nu}hh$ cross section.}
\setlength{\belowcaptionskip}{0pt}
\label{tab:r_coeffs_ee_nunuhh}
\vspace{-2mm}
\scriptsize
\setlength{\tabcolsep}{1pt}
\begin{tabular}{cc|ccccccccc}
\toprule
\textbf{Coeff.} & \shortstack{$\boldsymbol{\sqrt{s}}$\\[-1pt] [TeV]} & \textbf{Total} & \textbf{0--350} & \textbf{350--450} & \textbf{450--550} & \textbf{550--650} & \textbf{650--750} & \textbf{750--950} & \textbf{950--1350} & \textbf{1350}\,{\raisebox{0.2ex}{\tiny$\boldsymbol{+}$}} \\
\midrule
\multirow{13}{*}{$\sigma_{\mathrm{SM}}\,[\mathrm{ab}]$} & 0.35 & \pair{2.9}{-4}{2.8}{-7} & \pair{2.9}{-4}{2.8}{-7} & --- & --- & --- & --- & --- & --- & --- \\
~ & 0.50 & \pair{1.3}{-2}{1.2}{-5} & \pair{9.0}{-3}{8.8}{-6} & \pair{4.3}{-3}{5.1}{-6} & \pair{5.8}{-6}{9.4}{-9} & --- & --- & --- & --- & --- \\
~ & 0.75 & \pair{3.6}{-2}{2.5}{-5} & \pair{1.3}{-2}{8.5}{-6} & \pair{1.2}{-2}{8.4}{-6} & \pair{6.8}{-3}{5.9}{-6} & \pair{4.2}{-3}{4.1}{-6} & \pair{2.0}{-4}{2.8}{-7} & --- & --- & --- \\
~ & 1.00 & \pair{8.2}{-2}{6.3}{-5} & \pair{2.2}{-2}{2.5}{-5} & \pair{2.6}{-2}{3.4}{-5} & \pair{1.7}{-2}{1.4}{-5} & \pair{8.9}{-3}{7.1}{-6} & \pair{4.5}{-3}{4.0}{-6} & \pair{4.1}{-3}{5.0}{-6} & \pair{1.8}{-6}{2.7}{-9} & --- \\
~ & 2.00 & \pair{4.2}{-1}{4.7}{-4} & \pair{5.7}{-2}{7.4}{-5} & \pair{8.7}{-2}{9.5}{-5} & \pair{7.8}{-2}{9.7}{-5} & \pair{6.1}{-2}{4.1}{-5} & \pair{4.5}{-2}{5.7}{-5} & \pair{5.3}{-2}{4.6}{-5} & \pair{3.5}{-2}{4.1}{-5} & \pair{6.7}{-3}{5.4}{-6} \\
~ & 3.00 & \pair{8.4}{-1}{1.0}{-3} & \pair{8.3}{-2}{1.1}{-4} & \pair{1.3}{-1}{1.7}{-4} & \pair{1.3}{-1}{1.2}{-4} & \pair{1.1}{-1}{1.2}{-4} & \pair{8.8}{-2}{9.8}{-5} & \pair{1.2}{-1}{7.6}{-5} & \pair{1.2}{-1}{8.9}{-5} & \pair{5.6}{-2}{6.3}{-5} \\
~ & 4.00 & \pair{1.3}{0}{1.5}{-3} & \pair{1.0}{-1}{1.2}{-4} & \pair{1.6}{-1}{1.2}{-4} & \pair{1.7}{-1}{2.3}{-4} & \pair{1.5}{-1}{1.8}{-4} & \pair{1.2}{-1}{9.5}{-5} & \pair{1.9}{-1}{1.3}{-4} & \pair{2.1}{-1}{1.4}{-4} & \pair{1.6}{-1}{1.7}{-4} \\
\bottomrule
\end{tabular}
\end{minipage}%
}
\end{table*}

\begin{table*}[p]
\vspace*{-6mm}
\makebox[\textwidth][c]{%
  \hspace*{-1.4cm}%
  \begin{minipage}{\textwidth}
\centering
\caption{$\tilde{r}$-coefficients, SM cross sections, and their errors for the inclusive $e^+e^- \to Zhh$ cross section.}
\setlength{\belowcaptionskip}{0pt}
\label{tab:r_coeffs_ee_zhh}
\vspace{-2mm}
\scriptsize
\setlength{\tabcolsep}{1pt}
\begin{tabular}{cc|ccccccccc}
\toprule
\textbf{Coeff.} & \shortstack{$\boldsymbol{\sqrt{s}}$\\[-1pt] [TeV]} & \textbf{Total} & \textbf{0--350} & \textbf{350--450} & \textbf{450--550} & \textbf{550--650} & \textbf{650--750} & \textbf{750--950} & \textbf{950--1350} & \textbf{1350}\,{\raisebox{0.2ex}{\tiny$\boldsymbol{+}$}} \\
\midrule
\multirow{13}{*}{$\sigma_{\mathrm{SM}}\,[\mathrm{ab}]$} & 0.35 & \pair{3.5}{-3}{7.6}{-7} & \pair{3.5}{-3}{7.6}{-7} & --- & --- & --- & --- & --- & --- & --- \\
~ & 0.50 & \pair{1.6}{-1}{4.2}{-5} & \pair{1.0}{-1}{7.5}{-5} & \pair{5.6}{-2}{5.2}{-5} & --- & --- & --- & --- & --- & --- \\
~ & 0.75 & \pair{1.5}{-1}{5.0}{-5} & \pair{2.9}{-2}{2.5}{-5} & \pair{3.3}{-2}{4.1}{-5} & \pair{4.1}{-2}{4.5}{-5} & \pair{5.0}{-2}{5.4}{-5} & \pair{2.2}{-3}{5.5}{-6} & --- & --- & --- \\
~ & 1.00 & \pair{1.2}{-1}{5.2}{-5} & \pair{1.3}{-2}{3.3}{-6} & \pair{1.2}{-2}{1.3}{-5} & \pair{1.2}{-2}{2.0}{-5} & \pair{1.5}{-2}{2.6}{-5} & \pair{2.1}{-2}{2.5}{-5} & \pair{4.7}{-2}{4.4}{-5} & --- & --- \\
~ & 2.00 & \pair{5.5}{-2}{3.1}{-5} & \pair{2.2}{-3}{5.7}{-7} & \pair{1.5}{-3}{5.8}{-7} & \pair{1.1}{-3}{5.7}{-7} & \pair{9.4}{-4}{9.1}{-7} & \pair{9.3}{-4}{1.3}{-6} & \pair{2.2}{-3}{3.5}{-6} & \pair{7.9}{-3}{1.1}{-5} & \pair{3.8}{-2}{4.3}{-5} \\
~ & 3.00 & \pair{3.3}{-2}{2.2}{-5} & \pair{9.1}{-4}{2.4}{-7} & \pair{5.7}{-4}{1.9}{-7} & \pair{3.5}{-4}{1.2}{-7} & \pair{2.6}{-4}{1.2}{-7} & \pair{2.2}{-4}{1.7}{-7} & \pair{4.1}{-4}{4.5}{-7} & \pair{1.1}{-3}{1.7}{-6} & \pair{2.9}{-2}{2.8}{-5} \\
~ & 4.00 & \pair{2.2}{-2}{1.6}{-5} & \pair{5.0}{-4}{1.3}{-7} & \pair{3.0}{-4}{1.0}{-7} & \pair{1.7}{-4}{5.9}{-8} & \pair{1.2}{-4}{4.4}{-8} & \pair{9.0}{-5}{4.2}{-8} & \pair{1.5}{-4}{1.1}{-7} & \pair{3.0}{-4}{3.7}{-7} & \pair{2.1}{-2}{1.6}{-5} \\
\bottomrule
\end{tabular}
\end{minipage}%
}
\end{table*}

\begin{table*}[p]
\vspace*{-6mm}
\makebox[\textwidth][c]{%
  \hspace*{-2.6cm}%
  \begin{minipage}{\textwidth}
\centering
\caption{$\tilde{r}$-coefficients, SM cross sections, and their errors for the inclusive $e^-e^- \to e^-e^-hh$ cross section.}
\setlength{\belowcaptionskip}{0pt}
\label{tab:r_coeffs_ee_eehh_samesign}
\vspace{-2mm}
\scriptsize
\setlength{\tabcolsep}{1pt}
\begin{tabular}{cc|ccccccccc}
\toprule
\textbf{Coeff.} & \shortstack{$\boldsymbol{\sqrt{s}}$\\[-1pt] [TeV]} & \textbf{Total} & \textbf{0--350} & \textbf{350--450} & \textbf{450--550} & \textbf{550--650} & \textbf{650--750} & \textbf{750--950} & \textbf{950--1350} & \textbf{1350}\,{\raisebox{0.2ex}{\tiny$\boldsymbol{+}$}} \\
\midrule
\multirow{13}{*}{$\sigma_{\mathrm{SM}}\,[\mathrm{ab}]$} & 0.35 & \pair{7.1}{-6}{5.8}{-9} & \pair{7.1}{-6}{5.8}{-9} & --- & --- & --- & --- & --- & --- & --- \\
~ & 0.50 & \pair{3.5}{-4}{2.6}{-7} & \pair{2.9}{-4}{2.2}{-7} & \pair{6.4}{-5}{4.8}{-8} & \pair{5.1}{-7}{3.8}{-10} & --- & --- & --- & --- & --- \\
~ & 0.75 & \pair{3.3}{-3}{3.3}{-6} & \pair{1.5}{-3}{1.9}{-6} & \pair{1.3}{-3}{1.4}{-6} & \pair{4.8}{-4}{5.5}{-7} & \pair{9.6}{-5}{1.1}{-7} & \pair{4.9}{-6}{5.0}{-9} & --- & --- & --- \\
~ & 1.00 & \pair{9.6}{-3}{1.1}{-5} & \pair{2.9}{-3}{3.4}{-6} & \pair{3.3}{-3}{4.1}{-6} & \pair{2.0}{-3}{3.3}{-6} & \pair{9.8}{-4}{1.5}{-6} & \pair{3.7}{-4}{4.6}{-7} & \pair{1.1}{-4}{1.1}{-7} & \pair{1.2}{-7}{1.2}{-10} & --- \\
~ & 2.00 & \pair{5.5}{-2}{8.3}{-5} & \pair{7.6}{-3}{9.9}{-6} & \pair{1.1}{-2}{1.5}{-5} & \pair{1.0}{-2}{1.3}{-5} & \pair{8.0}{-3}{1.1}{-5} & \pair{5.8}{-3}{7.6}{-6} & \pair{7.0}{-3}{9.8}{-6} & \pair{4.5}{-3}{6.3}{-6} & \pair{5.3}{-4}{6.9}{-7} \\
~ & 3.00 & \pair{1.1}{-1}{1.5}{-4} & \pair{1.1}{-2}{1.5}{-5} & \pair{1.7}{-2}{2.4}{-5} & \pair{1.7}{-2}{2.3}{-5} & \pair{1.4}{-2}{2.0}{-5} & \pair{1.2}{-2}{1.4}{-5} & \pair{1.6}{-2}{2.1}{-5} & \pair{1.6}{-2}{1.8}{-5} & \pair{7.2}{-3}{9.7}{-6} \\
~ & 4.00 & \pair{1.7}{-1}{2.3}{-4} & \pair{1.3}{-2}{1.6}{-5} & \pair{2.2}{-2}{3.1}{-5} & \pair{2.2}{-2}{2.9}{-5} & \pair{2.0}{-2}{2.4}{-5} & \pair{1.6}{-2}{2.1}{-5} & \pair{2.5}{-2}{3.4}{-5} & \pair{2.8}{-2}{3.6}{-5} & \pair{2.1}{-2}{2.6}{-5} \\
\bottomrule
\end{tabular}
\end{minipage}%
}
\end{table*}

\begingroup
\tiny
\setlength{\tabcolsep}{1.6pt}
\begin{table}[p]
\centering
\caption{Per tag event numbers after cuts and $r$-coefficients for each $m_{4b}$ bin. WBF events enter the $e$-tag category only through the leptonic $W\to e\nu$ and contribute negligibly to $W$ coupling measurements. As can be identified by the number of events, $e$-tagging and $j$-tagging has little effect on the final results, even for $\kappa_{Z2}$.
} \label{tab:r_coeffs_binned_allinone}
\fontsize{7.4}{7.5}\selectfont
\renewcommand{\arraystretch}{1.00}
\begin{tabular*}{\linewidth}{@{\extracolsep{\fill}}ll|cccccc|ccccc|ccccc@{}}
\hline
\multicolumn{2}{c|}{} & \multicolumn{6}{c|}{Untagged} & \multicolumn{5}{c|}{$e^{\pm}$-tag} & \multicolumn{5}{c}{$j$-tag} \\
\cline{3-8}\cline{9-13}\cline{14-18}\noalign{\vskip 1.4pt}
Coeff. & $m_{4b}$ [GeV] & $ep$\,R & $ep$\,F & $ee$\,R & $ee$\,F & $\gamma\gamma$ & $\mu\mu$ & $ep$\,R & $ep$\,F & $ee$\,R & $ee$\,F & $\gamma\gamma$ & $ep$\,R & $ep$\,F & $ee$\,R & $ee$\,F & $\gamma\gamma$ \\ \noalign{\vskip 0.8pt}\hline\noalign{\vskip 1.5pt}
$N_{\mathrm{SM}}$ & $[0,350)$ & $1.2{\times}10^{3}$ & $378$ & $56.9$ & $37.0$ & $90.9$ & $159$ & $46.2$ & $4.56$ & $1.44$ & $0.76$ & $3.19$ & $37.5$ & $5.54$ & $2.11$ & $0.96$ & $7.27$ \\
 & $[350,450)$ & $2.0{\times}10^{3}$ & $560$ & $94.5$ & $62.0$ & $159$ & $236$ & $48.4$ & $5.47$ & $1.75$ & $0.92$ & $3.78$ & $38.7$ & $5.93$ & $2.26$ & $1.03$ & $7.60$ \\
 & $[450,550)$ & $1.7{\times}10^{3}$ & $569$ & $80.7$ & $52.1$ & $136$ & $245$ & $37.8$ & $4.63$ & $1.46$ & $0.78$ & $3.16$ & $29.3$ & $4.67$ & $1.78$ & $0.82$ & $6.08$ \\
 & $[550,650)$ & $1.2{\times}10^{3}$ & $471$ & $63.0$ & $43.5$ & $99.4$ & $216$ & $28.9$ & $3.91$ & $1.25$ & $0.68$ & $2.53$ & $22.2$ & $3.68$ & $1.42$ & $0.66$ & $4.62$ \\
 & $[650,750)$ & $880$ & $365$ & $48.1$ & $34.7$ & $71.0$ & $181$ & $21.6$ & $3.09$ & $0.97$ & $0.54$ & $1.93$ & $16.8$ & $3.03$ & $1.19$ & $0.57$ & $3.80$ \\
 & $[750,950)$ & $1.1{\times}10^{3}$ & $492$ & $64.6$ & $48.1$ & $91.2$ & $234$ & $26.4$ & $4.28$ & $1.43$ & $0.81$ & $2.72$ & $22.1$ & $4.14$ & $1.64$ & $0.78$ & $5.29$ \\
 & $[950,1350)$ & $878$ & $446$ & $55.3$ & $43.1$ & $76.3$ & $241$ & $21.8$ & $4.17$ & $2.26$ & $1.93$ & $2.53$ & $20.0$ & $4.11$ & $1.65$ & $0.80$ & $5.34$ \\
 & $[1350,5000)$ & $401$ & $247$ & $23.6$ & $17.8$ & $38.1$ & $152$ & $9.38$ & $2.80$ & $1.14$ & $1.00$ & $1.40$ & $7.47$ & $1.84$ & $0.76$ & $0.38$ & $2.81$ \\
\noalign{\vskip 1.6pt}\hline\noalign{\vskip 2.0pt}
$r_{3}$ & $[0,350)$ & $-1.60$ & $-1.69$ & $-1.70$ & $-1.69$ & $-1.63$ & $-1.69$ & $-0.29$ & $-0.42$ & $0.03$ & $0.70$ & $-0.55$ & $0.10$ & $0.22$ & $0.28$ & $0.31$ & $0.08$ \\
 & $[350,450)$ & $-1.20$ & $-1.20$ & $-1.19$ & $-1.17$ & $-1.26$ & $-1.20$ & $-0.64$ & $-0.29$ & $0.19$ & $0.74$ & $-0.72$ & $-0.75$ & $-0.68$ & $-0.65$ & $-0.63$ & $-0.75$ \\
 & $[450,550)$ & $-0.88$ & $-0.88$ & $-0.87$ & $-0.86$ & $-0.92$ & $-0.88$ & $-0.54$ & $-0.22$ & $0.10$ & $0.49$ & $-0.57$ & $-0.72$ & $-0.68$ & $-0.66$ & $-0.65$ & $-0.73$ \\
 & $[550,650)$ & $-0.68$ & $-0.67$ & $-0.68$ & $-0.67$ & $-0.74$ & $-0.67$ & $-0.44$ & $-0.25$ & $-0.11$ & $0.06$ & $-0.46$ & $-0.60$ & $-0.58$ & $-0.57$ & $-0.56$ & $-0.62$ \\
 & $[650,750)$ & $-0.59$ & $-0.59$ & $-0.56$ & $-0.55$ & $-0.63$ & $-0.59$ & $-0.39$ & $-0.18$ & $-0.07$ & $0.07$ & $-0.39$ & $-0.50$ & $-0.48$ & $-0.47$ & $-0.46$ & $-0.52$ \\
 & $[750,950)$ & $-0.50$ & $-0.50$ & $-0.48$ & $-0.47$ & $-0.55$ & $-0.50$ & $-0.24$ & $-0.14$ & $-0.08$ & $\sim\!0$ & $-0.27$ & $-0.41$ & $-0.40$ & $-0.39$ & $-0.38$ & $-0.44$ \\
 & $[950,1350)$ & $-0.43$ & $-0.43$ & $-0.41$ & $-0.40$ & $-0.47$ & $-0.43$ & $-0.16$ & $-0.13$ & $-0.11$ & $-0.09$ & $-0.18$ & $-0.30$ & $-0.29$ & $-0.28$ & $-0.27$ & $-0.32$ \\
 & $[1350,5000)$ & $-0.30$ & $-0.30$ & $-0.26$ & $-0.26$ & $-0.33$ & $-0.30$ & $-0.11$ & $-0.08$ & $-0.08$ & $-0.08$ & $-0.11$ & $-0.20$ & $-0.18$ & $-0.18$ & $-0.17$ & $-0.20$ \\
$r_{3,3}$ & $[0,350)$ & $2.48$ & $2.42$ & $2.29$ & $2.10$ & $3.06$ & $2.37$ & $1.09$ & $1.60$ & $1.63$ & $1.32$ & $1.49$ & $4.97$ & $5.00$ & $5.09$ & $5.07$ & $5.15$ \\
 & $[350,450)$ & $0.95$ & $0.95$ & $0.85$ & $0.80$ & $1.11$ & $0.94$ & $0.60$ & $0.81$ & $0.88$ & $0.90$ & $0.72$ & $2.36$ & $2.37$ & $2.42$ & $2.41$ & $2.42$ \\
 & $[450,550)$ & $0.51$ & $0.49$ & $0.47$ & $0.45$ & $0.62$ & $0.49$ & $0.36$ & $0.58$ & $0.68$ & $0.78$ & $0.42$ & $1.45$ & $1.46$ & $1.48$ & $1.47$ & $1.50$ \\
 & $[550,650)$ & $0.35$ & $0.33$ & $0.30$ & $0.28$ & $0.43$ & $0.33$ & $0.23$ & $0.29$ & $0.30$ & $0.30$ & $0.26$ & $0.98$ & $0.99$ & $1.00$ & $1.00$ & $1.03$ \\
 & $[650,750)$ & $0.29$ & $0.27$ & $0.22$ & $0.20$ & $0.34$ & $0.27$ & $0.19$ & $0.22$ & $0.22$ & $0.22$ & $0.20$ & $0.72$ & $0.73$ & $0.74$ & $0.74$ & $0.77$ \\
 & $[750,950)$ & $0.22$ & $0.21$ & $0.18$ & $0.17$ & $0.27$ & $0.21$ & $0.11$ & $0.14$ & $0.14$ & $0.13$ & $0.13$ & $0.55$ & $0.55$ & $0.56$ & $0.56$ & $0.58$ \\
 & $[950,1350)$ & $0.18$ & $0.18$ & $0.14$ & $0.13$ & $0.21$ & $0.18$ & $0.06$ & $0.07$ & $0.06$ & $0.06$ & $0.07$ & $0.34$ & $0.35$ & $0.35$ & $0.35$ & $0.37$ \\
 & $[1350,5000)$ & $0.07$ & $0.07$ & $0.05$ & $0.05$ & $0.09$ & $0.07$ & $0.02$ & $0.02$ & $0.02$ & $0.02$ & $0.02$ & $0.15$ & $0.15$ & $0.14$ & $0.14$ & $0.15$ \\
\noalign{\vskip 1.6pt}\hline\noalign{\vskip 2.0pt}
$r_{W2}$ & $[0,350)$ & $-2.57$ & $-2.64$ & $-0.99$ & $-0.50$ & $-2.47$ & $-2.61$ & $-0.09$ & $0.08$ & $0.22$ & $0.29$ & $-0.11$ & $-0.08$ & $0.11$ & $0.22$ & $0.27$ & $-0.10$ \\
 & $[350,450)$ & $-3.38$ & $-3.31$ & $-1.12$ & $-0.56$ & $-3.32$ & $-3.25$ & $-2.33$ & $-2.12$ & $-1.97$ & $-1.81$ & $-2.34$ & $-2.33$ & $-2.13$ & $-2.05$ & $-1.98$ & $-2.34$ \\
 & $[450,550)$ & $-3.98$ & $-3.92$ & $-1.15$ & $-0.56$ & $-3.84$ & $-3.90$ & $-3.54$ & $-3.32$ & $-3.17$ & $-2.97$ & $-3.61$ & $-3.54$ & $-3.35$ & $-3.29$ & $-3.20$ & $-3.61$ \\
 & $[550,650)$ & $-4.51$ & $-4.37$ & $-1.18$ & $-0.58$ & $-4.42$ & $-4.31$ & $-4.35$ & $-4.15$ & $-3.99$ & $-3.76$ & $-4.52$ & $-4.35$ & $-4.19$ & $-4.13$ & $-4.04$ & $-4.52$ \\
 & $[650,750)$ & $-5.38$ & $-5.26$ & $-1.23$ & $-0.60$ & $-5.10$ & $-5.30$ & $-4.95$ & $-4.74$ & $-4.52$ & $-4.20$ & $-5.23$ & $-4.96$ & $-4.80$ & $-4.71$ & $-4.59$ & $-5.23$ \\
 & $[750,950)$ & $-6.66$ & $-6.57$ & $-1.38$ & $-0.67$ & $-6.32$ & $-6.48$ & $-5.88$ & $-5.74$ & $-5.48$ & $-5.11$ & $-6.43$ & $-5.88$ & $-5.80$ & $-5.70$ & $-5.56$ & $-6.43$ \\
 & $[950,1350)$ & $-9.94$ & $-9.87$ & $-1.66$ & $-0.81$ & $-9.20$ & $-9.84$ & $-7.42$ & $-7.18$ & $-6.74$ & $-6.06$ & $-8.18$ & $-7.42$ & $-7.31$ & $-7.16$ & $-6.89$ & $-8.18$ \\
 & $[1350,5000)$ & $-17.9$ & $-17.8$ & $-1.96$ & $-0.96$ & $-14.4$ & $-17.8$ & $-10.6$ & $-10.2$ & $-9.65$ & $-8.92$ & $-11.5$ & $-10.6$ & $-10.2$ & $-9.97$ & $-9.60$ & $-11.4$ \\
$r_{W2,W2}$ & $[0,350)$ & $6.19$ & $5.94$ & $2.68$ & $1.33$ & $7.45$ & $5.79$ & $13.1$ & $13.3$ & $13.4$ & $13.2$ & $13.8$ & $13.2$ & $13.6$ & $13.9$ & $13.9$ & $13.9$ \\
 & $[350,450)$ & $8.39$ & $8.32$ & $2.93$ & $1.44$ & $9.17$ & $8.19$ & $21.3$ & $21.3$ & $21.5$ & $21.2$ & $21.9$ & $21.4$ & $21.6$ & $22.1$ & $22.0$ & $22.0$ \\
 & $[450,550)$ & $11.7$ & $11.1$ & $3.80$ & $1.84$ & $13.5$ & $10.9$ & $35.0$ & $34.8$ & $35.1$ & $34.6$ & $36.2$ & $35.1$ & $35.4$ & $36.0$ & $35.9$ & $36.4$ \\
 & $[550,650)$ & $18.0$ & $17.1$ & $5.21$ & $2.51$ & $20.4$ & $16.9$ & $51.7$ & $51.6$ & $51.7$ & $51.0$ & $54.6$ & $51.8$ & $52.3$ & $53.0$ & $52.9$ & $54.7$ \\
 & $[650,750)$ & $27.0$ & $25.6$ & $6.90$ & $3.31$ & $29.8$ & $25.7$ & $72.1$ & $72.4$ & $72.2$ & $71.1$ & $77.0$ & $72.2$ & $73.2$ & $73.7$ & $73.5$ & $77.1$ \\
 & $[750,950)$ & $44.1$ & $41.2$ & $10.8$ & $5.19$ & $51.4$ & $40.1$ & $117$ & $119$ & $120$ & $120$ & $126$ & $117$ & $120$ & $120$ & $121$ & $126$ \\
 & $[950,1350)$ & $111$ & $109$ & $20.6$ & $9.95$ & $118$ & $109$ & $218$ & $227$ & $228$ & $231$ & $240$ & $218$ & $228$ & $229$ & $231$ & $240$ \\
 & $[1350,5000)$ & $240$ & $239$ & $29.7$ & $14.5$ & $230$ & $238$ & $461$ & $503$ & $521$ & $547$ & $525$ & $460$ & $496$ & $503$ & $517$ & $525$ \\
$r_{W2,3}$ & $[0,350)$ & $7.43$ & $7.05$ & $3.21$ & $1.59$ & $8.93$ & $6.87$ & $15.7$ & $15.7$ & $15.9$ & $15.6$ & $16.4$ & $15.8$ & $16.0$ & $16.4$ & $16.3$ & $16.5$ \\
 & $[350,450)$ & $5.39$ & $5.26$ & $1.90$ & $0.93$ & $5.96$ & $5.17$ & $13.9$ & $13.8$ & $14.0$ & $13.7$ & $14.3$ & $14.0$ & $14.1$ & $14.4$ & $14.3$ & $14.3$ \\
 & $[450,550)$ & $4.68$ & $4.39$ & $1.52$ & $0.73$ & $5.41$ & $4.32$ & $14.1$ & $14.0$ & $14.0$ & $13.8$ & $14.5$ & $14.1$ & $14.2$ & $14.4$ & $14.4$ & $14.6$ \\
 & $[550,650)$ & $4.81$ & $4.51$ & $1.41$ & $0.68$ & $5.53$ & $4.44$ & $14.1$ & $14.0$ & $14.1$ & $13.9$ & $14.8$ & $14.1$ & $14.2$ & $14.4$ & $14.4$ & $14.9$ \\
 & $[650,750)$ & $5.32$ & $5.04$ & $1.36$ & $0.65$ & $5.87$ & $5.05$ & $14.3$ & $14.4$ & $14.3$ & $14.1$ & $15.3$ & $14.3$ & $14.5$ & $14.6$ & $14.6$ & $15.3$ \\
 & $[750,950)$ & $5.94$ & $5.54$ & $1.44$ & $0.69$ & $6.86$ & $5.41$ & $15.8$ & $16.0$ & $16.0$ & $16.0$ & $16.9$ & $15.8$ & $16.1$ & $16.1$ & $16.2$ & $16.9$ \\
 & $[950,1350)$ & $8.35$ & $8.18$ & $1.56$ & $0.75$ & $8.88$ & $8.14$ & $16.8$ & $17.4$ & $17.4$ & $17.6$ & $18.3$ & $16.8$ & $17.4$ & $17.5$ & $17.6$ & $18.3$ \\
 & $[1350,5000)$ & $7.29$ & $7.11$ & $0.95$ & $0.46$ & $7.30$ & $7.07$ & $15.5$ & $15.6$ & $15.6$ & $15.5$ & $16.1$ & $15.5$ & $15.5$ & $15.5$ & $15.5$ & $16.1$ \\
\noalign{\vskip 1.6pt}\hline\noalign{\vskip 2.0pt}
$r_{Z2}$ & $[0,350)$ & $-0.20$ & $-0.32$ & $-1.96$ & $-2.44$ & $-0.36$ & $-0.35$ & $-0.57$ & $-0.78$ & $-0.03$ & $1.09$ & $-1.01$ & --- & --- & --- & --- & --- \\
 & $[350,450)$ & $-0.28$ & $-0.40$ & $-2.53$ & $-3.05$ & $-0.52$ & $-0.44$ & $-1.96$ & $-0.89$ & $0.55$ & $2.24$ & $-2.22$ & --- & --- & --- & --- & --- \\
 & $[450,550)$ & $-0.36$ & $-0.47$ & $-3.17$ & $-3.70$ & $-0.71$ & $-0.49$ & $-2.65$ & $-1.13$ & $0.36$ & $2.22$ & $-2.80$ & --- & --- & --- & --- & --- \\
 & $[550,650)$ & $-0.44$ & $-0.56$ & $-3.75$ & $-4.28$ & $-0.93$ & $-0.59$ & $-3.15$ & $-1.93$ & $-0.95$ & $0.17$ & $-3.36$ & --- & --- & --- & --- & --- \\
 & $[650,750)$ & $-0.55$ & $-0.66$ & $-4.37$ & $-4.89$ & $-1.19$ & $-0.61$ & $-3.85$ & $-1.95$ & $-0.85$ & $0.45$ & $-3.86$ & --- & --- & --- & --- & --- \\
 & $[750,950)$ & $-0.73$ & $-0.85$ & $-5.74$ & $-6.32$ & $-1.68$ & $-0.91$ & $-3.49$ & $-2.14$ & $-1.31$ & $-0.21$ & $-3.90$ & --- & --- & --- & --- & --- \\
 & $[950,1350)$ & $-1.22$ & $-1.37$ & $-8.92$ & $-9.62$ & $-3.01$ & $-1.37$ & $-3.89$ & $-3.38$ & $-3.06$ & $-2.47$ & $-4.59$ & --- & --- & --- & --- & --- \\
 & $[1350,5000)$ & $-2.23$ & $-2.36$ & $-15.5$ & $-16.3$ & $-6.46$ & $-2.33$ & $-6.57$ & $-5.85$ & $-5.77$ & $-5.60$ & $-6.77$ & --- & --- & --- & --- & --- \\
$r_{Z2,Z2}$ & $[0,350)$ & $0.34$ & $0.58$ & $3.52$ & $4.40$ & $0.63$ & $0.63$ & $2.78$ & $4.19$ & $4.38$ & $3.60$ & $3.87$ & --- & --- & --- & --- & --- \\
 & $[350,450)$ & $0.54$ & $0.83$ & $5.32$ & $6.43$ & $1.05$ & $0.95$ & $5.41$ & $7.38$ & $8.05$ & $8.16$ & $6.57$ & --- & --- & --- & --- & --- \\
 & $[450,550)$ & $0.84$ & $1.15$ & $7.69$ & $9.01$ & $1.70$ & $1.22$ & $8.70$ & $14.2$ & $17.1$ & $19.6$ & $10.3$ & --- & --- & --- & --- & --- \\
 & $[550,650)$ & $1.29$ & $1.65$ & $11.1$ & $12.7$ & $2.72$ & $1.76$ & $12.2$ & $16.0$ & $16.6$ & $16.9$ & $13.9$ & --- & --- & --- & --- & --- \\
 & $[650,750)$ & $1.94$ & $2.34$ & $15.5$ & $17.3$ & $4.20$ & $2.18$ & $18.8$ & $21.0$ & $20.7$ & $20.2$ & $19.8$ & --- & --- & --- & --- & --- \\
 & $[750,950)$ & $3.76$ & $4.45$ & $29.9$ & $33.0$ & $8.72$ & $4.81$ & $24.2$ & $30.3$ & $30.0$ & $28.7$ & $28.1$ & --- & --- & --- & --- & --- \\
 & $[950,1350)$ & $9.89$ & $11.2$ & $73.0$ & $78.7$ & $24.5$ & $11.3$ & $39.5$ & $48.5$ & $48.5$ & $46.7$ & $46.7$ & --- & --- & --- & --- & --- \\
 & $[1350,5000)$ & $22.6$ & $24.0$ & $158$ & $165$ & $65.4$ & $23.7$ & $70.0$ & $65.8$ & $65.2$ & $63.9$ & $71.7$ & --- & --- & --- & --- & --- \\
$r_{Z2,3}$ & $[0,350)$ & $0.40$ & $0.67$ & $4.13$ & $5.16$ & $0.74$ & $0.74$ & $3.39$ & $5.03$ & $5.20$ & $4.26$ & $4.67$ & --- & --- & --- & --- & --- \\
 & $[350,450)$ & $0.34$ & $0.52$ & $3.31$ & $4.00$ & $0.65$ & $0.59$ & $3.55$ & $4.82$ & $5.25$ & $5.34$ & $4.29$ & --- & --- & --- & --- & --- \\
 & $[450,550)$ & $0.34$ & $0.46$ & $3.09$ & $3.62$ & $0.68$ & $0.49$ & $3.50$ & $5.66$ & $6.75$ & $7.73$ & $4.14$ & --- & --- & --- & --- & --- \\
 & $[550,650)$ & $0.34$ & $0.44$ & $2.95$ & $3.37$ & $0.72$ & $0.47$ & $3.35$ & $4.26$ & $4.39$ & $4.41$ & $3.79$ & --- & --- & --- & --- & --- \\
 & $[650,750)$ & $0.38$ & $0.46$ & $3.03$ & $3.40$ & $0.83$ & $0.43$ & $3.75$ & $4.25$ & $4.22$ & $4.16$ & $3.96$ & --- & --- & --- & --- & --- \\
 & $[750,950)$ & $0.50$ & $0.59$ & $3.96$ & $4.36$ & $1.15$ & $0.64$ & $3.28$ & $4.08$ & $4.04$ & $3.87$ & $3.79$ & --- & --- & --- & --- & --- \\
 & $[950,1350)$ & $0.74$ & $0.84$ & $5.43$ & $5.86$ & $1.82$ & $0.84$ & $3.05$ & $3.53$ & $3.47$ & $3.30$ & $3.56$ & --- & --- & --- & --- & --- \\
 & $[1350,5000)$ & $0.68$ & $0.71$ & $4.69$ & $4.92$ & $1.96$ & $0.70$ & $2.28$ & $1.81$ & $1.76$ & $1.70$ & $2.19$ & --- & --- & --- & --- & --- \\
\end{tabular*}
\end{table}
\endgroup

\section{Analytic Cross Section Formula \label{app:vbf_amplitudes}}

In this appendix, we provide analytic insight into the dependence of the relevant cross sections on the $\kappa$ coefficients. These expressions are intended to clarify the structure of the coupling dependence and to complement the full numerical computation used in the analysis.

\subsection{Di-Higgs Vector Boson Fusion}

Here, we review the scattering of longitudinal \(W\) and \(Z\) bosons into \(hh\) and extract the leading \(\sqrt{s}\) dependence from the analytic calculation of di-Higgs VBF in \cite{Dobrovolskaya:1990kx}. This provides insight into the high-energy behavior and can be used as a cross-check of our MadGraph results (including our $\tilde{r}$- coefficients) when running the effective vector boson approximation.

 First, we establish our conventions. We define $p_{1,2}$ for incoming momenta; $k_{1,2}$ are outgoing momenta; $q = p_1 - k_1$; $q' = p_1 - k_2$; and $\varepsilon_i$ are the longitudinal polarization vectors. Then, di-Higgs via $V_L \, V_L \rightarrow h \, h$ has four diagrams: $s,t,u$-channel and contact interactions. They are
\begin{align}
    \mathcal{A}_s &= \frac{g_{hVV} \, g_{hhh}}{s - \MH^2} \cdot (\varepsilon_1 \cdot \varepsilon_2),
    \\
    \mathcal{A}_t &= \frac{g_{hVV}^2 }{t - \MV^2} \cdot \left[(\varepsilon_1 \cdot \varepsilon_2) - \frac{(q\cdot \varepsilon_1)(q\cdot \varepsilon_2)}{\MV^2}\right],
    \\
    \mathcal{A}_u &= \frac{g_{hVV}^2 }{u - \MV^2} \cdot \left[(\varepsilon_1 \cdot \varepsilon_2) - \frac{(q'\cdot \varepsilon_1)(q'\cdot \varepsilon_2)}{\MV^2}\right],
    \\
    \mathcal{A}_c &= g_{hhVV}(\varepsilon_1 \cdot \varepsilon_2).
\end{align}

In terms of the $\kappa$ framework defined in \cref{eq:lagrangiankappa}, the couplings are given by
\begin{align}
  g_{hVV} =  2 \kappa_V m_V^2/v, \quad g_{hhVV} = 2 \kappa_{V2} m_V^2/v^2, \quad
g_{hhh} = 3 \kappa_3 m_h^2/v
\end{align}
The cross section is 
\begin{align}
\sigma =&\frac{\delta^2 s}{32\pi v^4}+\frac{m_V^2}{16\pi v^4}  \left[4\kappa_V^4-\delta^2 (\xi+1) - \delta\, \kappa_V \left(-3\kappa_3 \xi + 2\kappa_V (\xi-1) + 4\kappa_V \log\Big(\frac{s}{m_V^2}\Big)\right)\right]\nonumber \\
&+ \frac{m_V^4}{32\pi v^4}\frac{1}{s} \left[\delta^2 \left(-2\xi^2 + 4\xi + 2\right) + 2\delta\, \kappa_V \left(-3\kappa_3 \xi (\xi+2) + 4\kappa_V \left(2\xi^2 - 3\xi + 6\right) - 2\kappa_V (\xi-2)^2 \log\Big(\frac{s}{m_V^2}\Big)\right) \right. \nonumber \\
&\left. + \kappa_V^2 \left(9\kappa_3^2 \xi^2 - 12\kappa_3 \kappa_V (\xi-1)\xi + 4\kappa_V^2 \left(\xi^2 - 6\xi + 5\right) - 8\kappa_V \xi (3\kappa_3 - 2\kappa_V) \log\Big(\frac{s}{m_V^2}\Big)\right)\right] \nonumber\\
&+\mathcal{O}\left(\frac{m_V^4}{s^2}\log\Big(\frac{s}{m_V^2}\Big)\right) 
\end{align}
where \(\delta \equiv \kappa_{V2}-\kappa_V^2\) and \(\xi\equiv m_h^2/m_V^2\), and terms of higher order in \(m_V^2/s\) have been neglected. Here we see the expected growth linear in \(s\), proportional to \(\delta\), when a mismatch between \(\kappa_V\) and \(\kappa_{V2}\) spoils the Standard Model cancellation that ensures perturbative unitarity.
In the limit \(\delta\to 0\), this leading contribution, together with several subleading terms, vanishes, and the cross section simplifies to
\begin{align}
\lim_{\delta \to 0}\sigma =&\frac{m_V^2\kappa_V^4}{4\pi v^4}  
+ \frac{m_V^4\kappa_V^2}{32\pi v^4}\frac{1}{s} \left[ 9\kappa_3^2 \xi^2 - 12\kappa_3 \kappa_V (\xi-1)\xi + 4\kappa_V^2 \left(\xi^2 - 6\xi + 5\right) - 8\kappa_V \xi (3\kappa_3 - 2\kappa_V) \log\Big(\frac{s}{m_V^2}\Big)\right] \\
&+\mathcal{O}\left(\frac{m_V^4}{s^2}\log\Big(\frac{s}{m_V^2}\Big)\right) \nonumber.
\end{align}

\subsection{Di-Higgs Photon Fusion: \texorpdfstring{$\gamma\gamma \to hh$}{gamma gamma -> hh}}
\label{app:ggHH}
For completeness, we summarize the implementation of the one-loop helicity amplitudes used to compute the  $\gamma\gamma \to hh$ cross section. The $\gamma\gamma \to hh$ amplitude depends on three coupling modifiers $(\kappa_3, \kappa_W, \kappa_{W2})$
through top-quark and $W$-boson sectors~\cite{Jikia:1992mt,Belusevic:2004pz,Bharucha:2020bhy}.

The SM helicity amplitudes were implemented from the analytic expressions of Ref.~\cite{Jikia:1992mt}. The Passarino--Veltman scalar integrals required by these amplitudes are evaluated
numerically with \textsc{LoopTools}~\cite{Hahn:1998yk}.
The implementation was validated by comparing the SM total cross section against \texttt{MadGraph5\_aMC@NLO}~\cite{Alwall:2014hca}, finding agreement within numerical integration uncertainties.
The parametrization, including the modified quartic gauge-Higgs coupling
$\kappa_{W2}$, follows Appendix~A of Ref.~\cite{Bharucha:2020bhy}.
The helicity amplitudes then take the form
\begin{align}
  \mathcal{M}_{++} =& \, B_{++}^t
    +\kappa_3\, T_t
    + \kappa_W^2\, B_{++}^W
    + \kappa_W\kappa_3\, T_W + \kappa_{W2}\, Q_W ,
  \label{eq:Mpp_ggHH}\\
  \mathcal{M}_{+-} =&\, B_{+-}^t
    + \kappa_W^2\, B_{+-}^W ,
  \label{eq:Mpm_ggHH}
\end{align}
where \(T_{t,W}\) denote triangle form factors, \(B_{\pm\pm}^{t,W}\) are box contributions,
and \(Q_W\) denotes the contribution proportional to the
quartic \(W^+W^-hh\) coupling.

Equation~\eqref{eq:Mpm_ggHH} shows that
\(\mathcal M_{+-}\) is independent of
\(\kappa_3\) and \(\kappa_{W2}\).
Consequently, for the scenario considered in the main text with
\(\kappa_W=1\), all sensitivity to the Higgs self-coupling originates
from \(\mathcal M_{++}\).

Since the amplitudes depend at most linearly on $\kappa_3$ and $\kappa_{W2}$, and quadratically on $\kappa_{W}$, the squared matrix element is necessarily a finite polynomial in these couplings.
Squaring Eqs.~\eqref{eq:Mpp_ggHH}--\eqref{eq:Mpm_ggHH}, the independent monomials are
\begin{align}
\bigl\{1,\,\kappa_3,\,\kappa_W^2,\,\kappa_3\kappa_W,\,\kappa_{W2},\,
\kappa_3^2,\,\kappa_3\kappa_W^2,\,\kappa_3^2\kappa_W,\,
\kappa_3\kappa_{W2}, 
\,\kappa_W^4,\,\kappa_3\kappa_W^3,\,
\kappa_W^2\kappa_{W2},\,\kappa_3^2\kappa_W^2,\,
\kappa_3\kappa_W\kappa_{W2},\,\kappa_{W2}^2\bigr\},
\end{align}
a 15-dimensional basis.
Consequently, evaluating the cross section at 15 independent benchmark  points at each center-of-mass energy uniquely determines the polynomial coefficients, allowing the cross section to be reconstructed exactly for arbitrary coupling values. We used the following SM inputs:
$m_h = 125.0\,\text{GeV}$, $m_t = 173.3\,\text{GeV}$, $m_W = 80.419\,\text{GeV}$,
$\alpha_{\rm em}^{-1} = 128.93$.

\end{document}